\documentclass[manuscript,nonacm]{acmart}

\usepackage{wrapfig}
\usepackage{colortbl}
\usepackage{soul}

\AtBeginDocument{%
  \providecommand\BibTeX{{%
    \normalfont B\kern-0.5em{\scshape i\kern-0.25em b}\kern-0.8em\TeX}}}

\setcopyright{none}
\copyrightyear{2024}
\acmYear{2024}
\acmDOI{XXXXXXX.XXXXXXX}

\acmConference[Working Paper '24]{Working Paper}{January}{2024}
\acmPrice{15.00}
\acmISBN{978-1-4503-XXXX-X/18/06}

\begin{document}

\title{Metaverse Perspectives from Japan: A Participatory Speculative Design Case Study}

\author{Michel Hohendanner}
\authornote{Both authors contributed equally to this research.}
\authornote{Also affiliated at Department of Computer Science, Technical University of Munich}
\email{michelhohendanner@gmail.com}
\orcid{0000-0003-1560-9655}

\affiliation{%
  \institution{Munich Center for Digital Sciences and AI, Munich University of Applied Sciences}
  \streetaddress{Lothstraße 34}
  \city{Munich}
  \country{Germany}
  \postcode{80335}
}

\author{Chiara Ullstein}
\authornotemark[1]
\email{chiara.ullstein@tum.de}
\orcid{0000-0002-4834-4537}
\affiliation{%
  \institution{Department of Computer Science, Technical University of Munich}
  \streetaddress{Boltzmannstraße 3}
  \city{Munich}
  \country{Germany}}

\author{Dohjin Miyamoto}
\email{miyamotoneuro@gmail.com}
\orcid{0009-0007-2376-5160}
\affiliation{%
  \institution{Tokyo University}
  \city{Tokyo}
  \country{Japan}
}

\author{Emma Fukuwatari Huffman}
\email{efhuffman@gmail.com}
\orcid{0009-0004-7805-1032}
\affiliation{%
  \institution{Kyoto Design Lab, Kyoto Institute of Technology}
  \city{Kyoto}
  \country{Japan}
}

\author{Gudrun Socher}
\email{gudrun.socher@hm.edu}
\orcid{0000-0001-7211-7724}
\affiliation{%
  \institution{Munich Center for Digital Sciences and AI, Munich University of Applied Sciences}
  \streetaddress{Lothstraße 34}
  \city{Munich}
  \country{Germany}
  \postcode{80335}
}

\author{Jens Grossklags}
\email{jens.grossklags@tum.de}
\orcid{0000-0003-1093-1282}
\affiliation{%
  \institution{Department of Computer Science, Technical University of Munich}
  \streetaddress{Boltzmannstraße 3}
  \city{Munich}
  \country{Germany}}

  \author{Hirotaka Osawa}
\email{osawa.a3@keio.jp}
\orcid{0000-0001-5779-8437}
\affiliation{%
  \institution{Faculty of Science and Technology, Keio University}
  \city{Yokohama}
  \country{Japan}}

\renewcommand{\shortauthors}{Hohendanner and Ullstein, et al.}

\begin{abstract}
  Currently, the development of the metaverse lies in the hands of industry. Citizens have little influence on this process. Instead, to do justice to the pluralism of (digital) societies, we should strive for an open discourse including many different perspectives on the metaverse and its core technologies such as AI. We utilize a participatory speculative design (PSD) approach to explore Japanese citizens’ perspectives on future metaverse societies, as well as social and ethical implications. Our contributions are twofold. Firstly, we demonstrate the effectiveness of PSD in engaging citizens in critical discourse on emerging technologies like the metaverse, offering our workshop framework as a methodological contribution. Secondly, we identify key themes from participants' perspectives, providing insights for culturally sensitive design and development of virtual environments. Our analysis shows that participants imagine the metaverse to have the potential to solve a variety of societal issues; for example, breaking down barriers of physical environments for communication, social interaction, crisis preparation, and political participation, or tackling identity-related issues. Regarding future metaverse societies, participants’ imaginations raise critical questions about human-AI relations, technical solutionism, politics and technology, globalization and local cultures, and immersive technologies. We discuss implications and contribute to expanding conversations on metaverse developments.
\end{abstract}




\maketitle

\section{Introduction}

\textit{“The promise of metaverse is limitless, and basically anything that we now do in a physical realm can be done in a metaverse, at least in theory [...], but we should not be naïve in considering the underlying risk and potential harm for users.”} \cite[p.14]{Dwivedi2022Metaverse}

There has been diverse criticism on the recently revived idea of the metaverse such as being yet another example of inequalities-reproducing futuristic world-making \cite{Bell2022metaverse}. Despite this criticism, the metaverse, as questionable “promise of tomorrow” \cite[p.211]{Knox2022Metaverse}, represents a good use case for directing critical thought towards exploring sociotechnical developments and underlying ethical issues for the future. This appears in particular of importance in a time where the metaverse has recently gained new momentum driven by Meta’s (formerly Facebook) decision, among other companies like Microsoft, Epic Games, Tencent or Niantic, to focus on it as their future development strategy \cite{Gorichanaz2022Being, Meta2021Facebook, Xu2022Metaverse:}. As there is still little academic discourse to scientifically guide or accompany related investments \cite{Duan2021Metaverse}, the sovereignty of interpretation of future development currently lies in the hands of industry. 

Building on narrative conceptual influences like the novels \textit{True Names} and \textit{Neuromancer}, the metaverse as terminology was first introduced in 1992 in the science fiction novel \textit{Snow Crash} \cite{Dionisio20133D}. It combines the words meta, i.e., beyond, and verse as in universe, referring to a computer-generated universe beyond the physical world \cite{Dionisio20133D}. More specifically, the term describes an “integrated network of 3D virtual worlds [...] that constitutes a compelling alternative realm for human sociocultural interaction” \cite[p.2]{Dionisio20133D}. More broadly, the term can also stand for ``the seamless convergence of our physical and digital lives'' with  ``a set of interoperable virtual spaces where we can work, play, learn, relax, socialise, communicate, interact, transact, and own digital assets'' being the core aspect of this convergence \cite[p.60]{DelphiMetaverse_2023}.

The metaverse -- as upcoming social imaginary \cite{Gorichanaz2022Being} -- represents a vision that could deeply affect social structures of future societies, while offering the potential to improve “the access and experience of several services in sectors such as education, healthcare, and culture” \cite[p.14]{Dwivedi2022Metaverse}. However, given the speed of metaverse-related technology innovation and the scale of industry investments, stakeholders have issued urgent calls to close the gap between academia and industry-driven perspectives on metaverse development \cite{Duan2021Metaverse, Xu2022Metaverse:}, shift the focus from economic potential to human factors and social implications of a prospective metaverse, and enhance a diversified discourse \cite{Duan2021Metaverse, Dwivedi2022Metaverse, Xu2022Metaverse:}. Our work addresses these goals and, in particular, aims for a diversified discourse by including citizens as everyday experts of their individual (local) realities to provide a case study of what a prosocial mode of engagement can look like. 

Following this approach, this research study investigates citizens’ perspectives on the metaverse and associated societal developments through the lens of participatory speculative design (PSD) \cite{Farias2022Social}. We contrast our study to previous research that mainly focuses on the perception of metaverse technology’s capacities in specific domains, e.g., teachers' and students' perception in education, or on the perception of a specific aspect of the technology, e.g., the influence of body representation through avatars on body awareness. We explore how the overall idea of a prospective metaverse as a socio-technical system and its promise of innovation is perceived by citizens in an in-depth manner, without pre-framing a specific aspect or context of use. We pose the following research questions:
\begin{itemize}
    \item RQ1: How do citizens in Japan make sense of a prospective metaverse? Through posing this question, we investigate citizens' perceived problem spaces to which the technology can be applied and the envisioned capacities of the technology to tackle those problems. 
    \item RQ2: How do citizens in Japan make sense of a future in which a metaverse holds a central position? Through this question, we investigate citizens' perceptions of societal transformations in an imagined future where the metaverse moves to the foreground. Linked to this, we explore societal issues and value discussions imagined by citizens as a consequence of these transformations. 
\end{itemize}
We chose Japan as a use case for this investigation, rolling out a two-day PSD workshop with focus groups. Dionisio et al. \cite{Dionisio20133D} identify technical factors, institutional agendas and the interest of the public as key drivers for a viable metaverse. These are strongly pronounced in Japan, illustrated e.g., by the governmental agenda to reach Society 5.0 \cite{CabinetOfficeGovernmentofJapan2017Society} or the popularity of avatar culture \cite{Bredikhina2020Avatar} and VTubers \cite{Lu2021More}. We further introduce the utilized PSD toolkit, which guides workshop participants to imagine future metaverse societies and critically reflect on them by creating fictional narratives in the form of magazine articles. We find that participants imagine the metaverse as technology with the potential to solve a variety of societal issues such as overcoming barriers of physical environments for communication, social interaction, crisis preparation, and political participation, or tackling identity-related issues. Two out of four groups put a strong emphasis on the role of artificial intelligence (AI) within a prospective metaverse. Participants also raise critical questions about human-AI relations, technical solutionism, politics and technology, globalization and local cultures, and immersive technologies.

We contribute to computer-supported cooperative work (CSCW) and human-computer interaction (HCI) with an empirical study on a citizens-generated vision of the future of metaverses. 
Our contributions to the field are twofold:
\begin{enumerate}
    \item We demonstrate the efficacy of PSD in the field of HCI and CSCW in opening up discursive spaces to laypeople and provide our workshop framework as a methodological contribution (see toolkit in Appendix \ref{App:WSToolKIT}):
    \begin{enumerate}
        \item [a)] Our framework enabled participants to critically reflect on the metaverse and its potential benefits and risks, providing a unique way to engage with and understand citizen perceptions and expectations concerning emerging technologies.
        \item [b)] The framework exemplifies a decentralized application of SD techniques, fostering the inclusion of a wide array of locally-specific  contributions by citizens to socio-technical discourses.
    \end{enumerate}
    \item We present themes that are at the forefront of citizens' minds concerning the metaverse as a solution to future problem spaces as well as societal effects and risks that emerge from the technology, and discuss their implications for metaverse and AI design and development:
    \begin{enumerate}
        \item [a)] With the workshop case study, we add a deep nuanced understanding of how participants in Japan think about a future metaverse. We emphasize the need for culturally sensitive design approaches in collaborative virtual environments.
        \item [b)] We relate themes foregrounded by citizens to prior HCI and CSCW research. We report on the implications of identified themes, which could benefit in prioritizing design efforts, and identify important questions that future research should address.
    \end{enumerate}
\end{enumerate}
With both contributions, we intend to stress the importance of citizen-centric as well as local-cultural sensitive research, design, and development of a prospective metaverse.

This paper unfolds as follows: First, we outline conceptualizations and perceptions of the metaverse, present recent metaverse-related developments in Japan, and relate this case study to recent work in the field of research through design for CSCW and HCI. In Section 3, we outline how the use case was rolled out in Tokyo/Yokohama in mid-2022. In Section 4, we analyze underlying themes focusing on two different levels of sensemaking: sensemaking of the metaverse as emerging technology and sensemaking of prospective societies through the application of metaverse technology. We discuss results and outline implications in Section 5. Before concluding, we reflect on limitations of the study and provide an outlook for further research.

\section{BACKGROUND AND RELATED LITERATURE}

In this section, we introduce conceptualizations and perceptions of the metaverse that will later serve as a foundation for the analysis and discussion of the workshop results. We relate our research to existing studies in the fields of CSCW and HCI regarding  perceptions of the metaverse, as well as the application of speculative design approaches. Contextualizing our study in Japan, we outline recent metaverse-related developments in Japan and Japanese VR culture.

\subsection{From Approaching the Concept of the Metaverse to Sociotechnical Imaginaries and Public Perception}

AI has been identified to be one of the core technologies to drive the development of the metaverse \cite{Yang2023Human-Centric}. Dionisio et al. \cite{Dionisio20133D} identify four main areas for development that could enable the move from various independent virtual worlds to a single metaverse: “immersive realism, ubiquity of access and identity, interoperability, and scalability” \cite[p.1]{Dionisio20133D}.

Key challenges in metaverse development are not solely of a technical nature, but to a great extent depend also on social aspects. Duan et al. \cite{Duan2021Metaverse} introduce representative applications of a metaverse for social good: accessibility, e.g., in terms of global communication and the unrestricted attendance at social and cultural events; diversity, e.g., providing spaces of interaction and expression without physical limitations of space and distances; equality, e.g., by questioning social constructs like race and gender through the use of avatars; humanity, e.g., in the digital perseverance of cultural achievements like world-famous architecture. To discuss social implications of a potential AI-driven metaverse, we later draw on these principles in the discussion of the workshop results. This resonates with the pursuit towards a human-centric metaverse proposed by Yang et al. \cite{Yang2023Human-Centric}.

The metaverse can also be seen as constituting an upcoming social imaginary \cite{Gorichanaz2022Being}, which emerges “in the culture, as expressed through imagery, stories, mass media and the like” \cite[p.4]{Gorichanaz2022Being}. Social imaginaries related to digital technology, as in the case of the metaverse, are called sociotechnical imaginaries \cite{Jasanoff2015Dreamscapes}\footnote{The researchers have been following the developments around the \#metoosts initiative and express their support to all victims of abuse in this context.}. Social and sociotechnical imaginaries “shape our understanding of our current situation while also inspiring and guiding our actions into the future” \cite[p.4]{Gorichanaz2022Being}. Reaching the status of an applicable sociotechnical imaginary means that not only is metaverse development shaped by what humans imagine it to be, but so are its intersections with other aspects of future human social life. This also affects what a prospective society is imagined to look like. As the metaverse is increasingly imagined to be an important part of future societies, it becomes more likely that it will become part of them.

To the best knowledge of the authors, few peer-reviewed studies focusing on the public’s general perception of social and societal opportunities and risks of the metaverse exist. Some of these studies are preprints or published as blog posts and reports. A recent study found 63\% of participants from India to be concerned about a loss of the link to reality~\cite{Ananya2022Impact}. Despite concerns, two studies show that respondents from the US, UK, and India believed that people can benefit from possibilities of overcoming real-world limitations such as traveling \cite{Ananya2022Impact, YouGov2022Unlocking}. Some respondents (48\%-60\%) from the US and UK would be interested in entering the metaverse with an alter ego \cite{YouGov2022Unlocking}. Generally, the majority of English-speaking \cite{Hayawi2022Inevitable-Metaverse} and Turkish \cite{Akkuzukaya2022Sentiment} Twitter users showed positive sentiments. Citizens from Indonesia showed predominantly neutral sentiments \cite{Ahmad2022Sentimen}. Recently, studies have explored user's acceptance of the metaverse utilizing an (extended) technology acceptance model, for example, identifying perceived usefulness and perceived ease of use to positively influence attitudes toward using the metaverse \cite[e.g.,][]{park2021TAM, Wu2023TAM, Aburbeian2022Technology}. These studies provide quantitative accounts of peoples' perceptions of the metaverse, which inevitably limits the spectrum of potential responses from subjects. Our study complements these studies with an qualitative approach enabling participants to express their imaginations of a future metaverse, helping us to derive, on the one hand, use cases where the metaverse is perceived to be useful, and on the other hand, consequences that could emerge from depicted future metaverse societies.

Besides research on the general perception, previous research in the fields of HCI and CSCW has mainly focused on understanding how different stakeholders perceive the metaverse in various specific domains. For instance, in the context of \textit{education}, studies have concentrated on the perspectives of students and teachers regarding the use of the metaverse as an educational environment \cite{Paananen_Empathy_MetaverseCampus2023, ClasseStudentsTechAcc2023, CantoneLanguageLearningMetaverse2023, YunTeacherStudentMetaverse2022}. In the domain of \textit{remote work}, research has targeted employers or executive-level employees experienced in utilizing metaverse applications as workspaces \cite{ParkMetaverseWorkspace2023}. With a focus on \textit{virtual product design}, research investigated how designers perceive the metaverse for creating virtual product experiences \cite{CilizogluDesignersMetaverse2023}. In the field of \textit{accounting and financial services}, attention was given to auditors, particularly those working in public accounting firms, and their perception of the metaverse \cite{LindawatiAccountantMetaverse2023}. Research on \textit{child-related activities} targeted perceptions related to play and education in the metaverse, particularly from experts and parent communities \cite{LimChildMetaverse2023}. Investigating \textit{political processes}, prior work has explored how UK residents perceive the use of smart contracts in the metaverse for voting \cite{OppenlaenderDemocracyMetaverse2022}.

Additionally, some studies have delved into how certain aspects of the metaverse affect human perception. These include the influence of avatars on body awareness \cite{DoellingerAvatarBodyAware2023}, workgroup inclusion \cite{BuckAvatarGroupIncl2023} and immersiveness of full body interactions \cite{LamAvatarBodyInterac2022}. Others investigate the perceived impact of harmful metaverse environment design \cite{KouHarmfulMetaverse2023} or users' privacy-related behaviors in social VR applications \cite{SykownikPrivacyBehMetaverse2022}. Methods of inquiry are mainly (online) surveys \cite[e.g.,][]{Paananen_Empathy_MetaverseCampus2023, YunTeacherStudentMetaverse2022, LamAvatarBodyInterac2022, SykownikPrivacyBehMetaverse2022} and interviews \cite[e.g.,][]{ParkMetaverseWorkspace2023, CilizogluDesignersMetaverse2023, LindawatiAccountantMetaverse2023}, followed by user experience case studies \cite[e.g.,][]{DoellingerAvatarBodyAware2023, ClasseStudentsTechAcc2023, BuckAvatarGroupIncl2023}.

While these studies on specific stakeholders' perceptions of pre-defined metaverse or VR applications represent valuable contributions to the field, studies focusing on a broader perception of the metaverse as a socio-technical phenomenon seem to be underrepresented. To us, the latter perspectives seem especially valuable as an established future metaverse would have an impact on significant parts of future societies.

\subsection{Metaverse and VR Culture in Japan}

In the Japanese context, we observe only limited academic research efforts on the metaverse as a socio-technical system. In an industry study conducted in Japan, 77\% of all study participants who expressed (much) interest in the metaverse (24\%) indicated to look for experiences that are not possible in real life \cite{CrossMarketing2022}. In the same study, the three most common associations with the metaverse by Japanese participants were virtual space, VR/virtual reality, and Facebook~\cite{CrossMarketing2022}. 

Compared to the US and the UK, in Japan there has been a steadily higher interest in the term metaverse in Google searches \cite{Google2023Google} in the 12 months following Mark Zuckerberg’s announcement to rename Facebook into Meta and to focus on developing the metaverse \cite{Meta2021Facebook}. Even before the announcement, Japan was one of the leaders of VR development and usage. This can be observed on a macro level by looking at Japan’s official government vision of the realization of Society 5.0 \cite{CabinetOfficeGovernmentofJapan2017Society}, which stands for the creation of a society where cyberspace and the physical realm are closely intertwined for the purpose of solving social challenges \cite{CabinetOfficeGovernmentofJapan2017Society}. As part of this agenda, but also beyond, several investments were made and initiatives started in Japan. An example for governmental involvement represents funded research on cybernetic avatars as part of the Moonshot R\&D program \cite{CabinetOfficeGovernmentofJapan2022Moonshot}. Furthermore, the University of Tokyo has recently launched the metaverse School of Engineering as an academia-industry collaboration \cite{TheUniversityofTokyo2022Announcing}, and students attend entrance ceremonies \cite{Blaster2022Japanese} or job fairs in virtual spaces \cite{Ishiyama2023Metaverse}. Large Japanese corporations such as Sony aim to play a leading role in metaverse development \cite{Nussey2022Sony}. Summits and exhibits, amongst others organized by Meta “to promote the metaverse from Japan to the world” \cite{Yamagishi2022METAVERSE}, showcase the latest technologies \cite{MetaverseExpoTokyo2022Metaverse,MetaverseSummitJapan2022}.

These developments also draw upon Japan’s rich and well-established VR culture \cite{Bredikhina2020Avatar}, which likewise plays an important role in the advancement of VR technology in general. In the past, VR went through four development phases \cite{Bredikhina2020Avatar}. The current phase VR 4.0 stands for enabling a human ecosystem open for social and commercial networks \cite{Bredikhina2020Avatar}. In this context, the Japanese avatar-driven VR society is considered an important development factor, especially regarding the wide spread of VR interaction environments and industry structure \cite{Bredikhina2020Avatar}. One specific driver are VTubers that originated in Japan in 2016 \cite{Lu2021More}. VTubers are represented by animated virtual avatars and perform in recorded or live videos \cite{Lu2021More}. The level of popularity VTubers already reach is shown, for example, by the fact that in 2018 the Japanese National Tourism Organization selected Kizuna Ai, the avatar of the very first VTuber, as an ambassador for an international culture campaign \cite{JapanNationalTourismOrganizationNewYork2018JNTO}. Beyond VTubers’ presence on established video platforms like YouTube, the platforms VRChat, Virtual Chast, SHOWROOM, REALITY, Mirrativ and Cluster were identified as important drivers for the development of VR 4.0 \cite{Bredikhina2020Avatar}. These platforms provide important virtual spaces for interactions between audience and content creators including VTubers \cite{Bredikhina2020Avatar}. VRChat already hosted several pioneering virtual social events like Virtual Market, a series of exhibition and trade conventions focusing on comic, gaming and manga content \cite{Bredikhina2020Avatar}. A study conducted with 576 VRChat users in 2019 revealed that 41\% of users were inspired to try VRChat by watching VTuber content, while 30\% were invited by friends \cite{Shinbo2019}. 57\% of users indicated they were interested in VR communication and 64\% indicated to be mainly interested in avatar-driven social interactions \cite{Shinbo2019}. 60\% of users indicated buying avatars and having a preference to customize them \cite{Shinbo2019}. 
This interest in commercial activities was also observed in a recent study on metaverse perceptions among Japanese citizens \cite{CrossMarketing2022}. In this study, 24\% of participants showed a high interest in the metaverse. Within this subgroup, 68\% of participants expressed a high desire in shopping and working with cryptocurrency and NFTs in virtual space, 66\% indicated to be interested in participating in avatar communities in virtual space and interacting with others, and 77\% in virtual cutting-edge experiences that cannot be experienced in real life \cite{CrossMarketing2022}.

Summarizing, we primarily observe research on the perceptions regarding existing VR applications, which is an anticipated finding given their wide adoption in Japan. We are not aware of studies that provide citizens a space for discourse to explore the metaverse as a socio-technical system, and to elaborate on what a pervasive introduction would mean for society. From a methodology perspective, we have identified a study \cite{Trucchia5G2023} on the exploration of 5G in Japan that is compatible with our approach. Overall, the market and technology landscape distinguishes Japan from other countries, however, it does not allow the general conclusion that all Japanese citizens are more experienced with VR technologies and avatars. In any case, the existing conditions make Japan an appealing use case to explore notions of future metaverse societies and their implications.

\subsection{Sensemaking and Speculative Approaches in Research through Design for HCI}

This research applies a research through design (RtD) \cite{Frayling1994Research} approach, meaning it uses the act of designing to generate new knowledge \cite{Zimmerman2014Research}. Due to a shifting focus within third-wave HCI research and related knowledge production \cite{Höök2015Knowledge}, RtD is increasingly being applied in research and practice. The act of designing is subject to research into design \cite{Frayling1994Research} and has been described to integrate a \textit{problem-solving} dimension \cite{Simon1969sciences} as well as an interdependent \textit{sensemaking} dimension \cite{Manzini2015Design}. The dimension of problem-solving is related to the physical and biological world, where technical solutions for problems can be found regarding the form, utility, and function of design applications. Sensemaking, in turn, relates to the social world, where the desirability of a design solution can be assessed, i.e., meaning can be produced and cultural quality can be assessed. Therefore, design practice and its outcomes can be utilized to understand how people perceive their physical and social environment. Likewise, as design inherently connects social and technical systems \cite{Manzini2015Design}, underlying value systems that flow into design processes can be revealed. In this light, it has been contemplated that design research can be utilized for a critical exploration of contemporary challenges, “through a variety of practices, methods, and perspectives, including (but not limited to) Research through Design, Critical Design, Speculative Design and Participatory Design” \cite[p.2]{Lindley2022Communicating}. The following provides examples of how different modes of design speculation are applied as RtD in HCI and beyond.

Speculative design (SD) is a design practice that emerged from critical design and distinguishes itself from mainstream design by not necessarily having to follow a commercial logic, rather its value is rooted in the imaginative \cite{Dunne2013Speculative, Johannessen2019Speculative}. A key goal of SD is to stimulate the public to critically reflect and negotiate common norms and values regarding identified problem areas \cite{Johannessen2019Speculative}. The related approach design fiction (DF) \cite{Bleecker2009Design} describes the ``deliberate use of diegetic prototypes to suspend disbelief about change” \cite{Sterling2013Patently}. DF aims to bridge imagination and materialization “by crafting, modeling things and telling stories through objects” resulting in conversation pieces \cite[p.8]{Bleecker2009Design}. Similarly, SD uses so-called props \cite{Johannessen2019Speculative} to foster discourses. Props are digital or physical visual proposals for representations of speculated futures \cite{Johannessen2019Speculative} or artifacts “stolen” \cite{Lutz2020Future} from them. By responding to what-if questions regarding pre-identified future challenges \cite{Dunne2013Speculative}, SD props are provocative and shall trigger questions about how the future of social and societal living environments may look like \cite{Dunne2001Design,Mitrović2015introduction}. They can take different formats, for example a fictional product catalog \cite{Brown2016IKEA,NearFutureLaboratory2014TBD} or newspaper \cite{O’Shea2014Winning}, speculative product and service websites \cite{Hohendanner2021Designing_not_anonym, Hohendanner2023ReflectiveSpace}, imaginative forum questions or posts \cite{Wong2018When}, or scenes and software prototypes \cite{Gerber2018Participatory}. Besides objects, narratives are another mode of materialization of speculations \cite{johannessen2017young}. Narratives written by study participants have been used to explore environmental and energy consumption concerns \cite{Prost2015From} or romance and friendship during the pandemic \cite{Sharma2021From}. Other studies use prepared speculative narratives as props to elicit critical reflections from participants \cite{Schulte2016Homes}, for example, to catalyze meaningful conversations on AI recruitment processes between job applicants and HR recruiters \cite{Kaur2022Work}, to explore responses to fictional technologies \cite{Dalton2016Resistance}, or to elicit values regarding innovation in the identity space from marginalized groups \cite{Briggs2015Inclusive}.

While design speculations have been widely applied in the last decade, criticism and limitations have to be taken into account. Core of recent criticism is the question of who is involved in the speculation processes. Design speculations have been criticized to be elitist and patriarchal \cite{Light2021Collaborative,PradodeMartins2014Privilege,Tonkinwise2014How}; to suppress local cultural specifics if professional designers work without involving affected stakeholders \cite{Drazin2020Designte}; to address a limited spectrum of critique \cite{Farias2022Social, Sengers2021Speculation,Ward2021Practice}; or to be based on market logic \cite{Tonkinwise2014How}. Following this criticism, it can be observed that SD in particular is taking a participatory turn \cite{Farias2022Social}. PSD \cite{Baumann2018Participatory, Farias2022Social, Korsmeyer2019Learning} aims at democratizing design speculations by opening the speculation process to non-designers and decentralizing power systems to reach and include a wide public into the development process \cite{Korsmeyer2019Learning}, making way for decolonial PSD \cite[e.g.,][]{Bray2021Speculative, Khan2021Speculative, TranoLeary2019Who, Winchester2018Afrofuturism}, feminist approaches to PSD \cite[e.g.,][]{Sondergaard2018Intimate} and the empowerment of issue-specific or technology-specific groups \cite[e.g.,][]{Elsden2017On,Kaur2022Work,Prost2015From,Ventä-Olkkonen2021Nowhere}. 

\section{METHODOLOGY}

With our research, we intend to expand the discourse around a prospective metaverse by integrating citizens' perceptions on its development, to enhance the prospective metaverse's social sustainability and inclusiveness. This includes the goal of creating spaces for informed discourse among citizens and enabling processes to disseminate the diverse perspectives that surface through these discourses. The rationale behind these processes is to complement perspectives of developers, industry stakeholders and respective research communities with citizens’ perspectives as a key prospective user group.

In the previous section, we have established that most prior work focuses on the perception of metaverse technology’s capacities in specific domains, e.g., education, or on the perception of a specific aspect of the technology, e.g., body representation through avatars. Instead, we aim to document and analyze what comes to citizens’ minds when engaging in a discourse about a prospective metaverse as a socio-technical system without pre-framing a specific aspect or context of use. This approach brings into focus the perceived potential benefits and risks of the technology in relation to its anticipated social impact, and allows to highlight what citizens perceive as acceptable use or not. Furthermore, our approach allows citizens to illustrate how they imagine a prospective use of the technology.

This section highlights our research procedures with respect to this study's goals, purposes and applied research perspective. This is illustrated by adapting the Goal-Question-Metric (GQM) \cite{Wohlin2012GQM}: The main goal of this study, to foreground citizens' perceptions, can be split into two measurement goals when applying the GQM goal template \cite{Briand1996applyGQM}, detailed in Table~\ref{tab:GQM}: First, each goal is described through its object of study, its purpose, its focus or perspective, the applied viewpoint and the context of the study. Second, each goal is characterized through a set of questions. Third, the type of data that is expected to address each of these questions and the analysis procedures are outlined. 

In addition, we provide further details regarding the rationale of our workshop approach, the recruitment of participants, the workshop design, and the analysis approach of the collected data.

\begin{table}[hbt]
    \caption{Goals, purpose, and the perspective of the study applying GQM}
    \label{tab:GQM}
    \small
    \begin{tabular}{p{0.1\linewidth} p{0.41\linewidth} p{0.41\linewidth}}
    \toprule
        GQM\newline Category & Goal 1: Characterize Metaverse as Solution & Goal 2: Characterize Imagined Future Societies \\
    \midrule
        Measurement Goals & 
        - \underline{Analyze} citizens’ imaginations of prospective metaverse use cases \newline
        - \underline{for the purpose of} characterization\textsuperscript{1} \newline
        - \underline{with respect to} imagined \textit{solution competence} of a prospective metaverse \newline
        - \underline{from the viewpoint of} the researchers \newline
        - \underline{in the context of} a participatory speculative design workshop in Tokyo/Yokohama, Japan.
        & 
        - \underline{Analyze} citizens’ imaginations of prospective metaverse use cases \newline
        - \underline{for the purpose of} characterization\textsuperscript{1} \newline
        - \underline{with respect to} imagined \textit{social impact} of a prospective metaverse \newline
        - \underline{from the viewpoint of} the researchers \newline
        - \underline{in the context of} a participatory speculative design workshop in Tokyo/Yokohama, Japan. \\ 
    
    \arrayrulecolor{lightgray}\hline       
        Questions & 
        Q1.1: When considering future societies, what problem space(s) do citizens perceive that the metaverse could be a solution for? \newline 
        Q1.2: How do citizens envision the metaverse to tackle identified problems? 
        & 
        Q2.1: What transformations in society do citizens anticipate with the metaverse becoming central to a future society? \newline 
        Q2.2: Which societal issues are discussed as consequence of the metaverse holding a central societal position?
        \\
    \hline    
        Metrics & 
        For Q1.1: Deduction of problem spaces inherent in participants' narratives identified through qualitative analysis (actantial model and process coding; see Table \ref{tab:sensemaking_metaverse} col. 2). \newline 
        For Q1.2: Deduction of the solution competence of metaverse technology inherent in participants' narratives identified through qualitative analysis (actantial model and process coding; see Table \ref{tab:sensemaking_metaverse} col. 3).
        & 
        For Q2.1: Deduction of anticipated societal transformations inherent in participants' narratives identified through qualitative analysis (actantial model, process coding and values coding; see Table \ref{tab:sensemaking_future} col. 2).
         \newline 
        For Q2.2: Deduction of consequential societal problem spaces and value discussions emerging from participants' narratives identified through qualitative analysis (actantial model and values coding; see Table \ref{tab:sensemaking_future} col. 3).
        \\
    \arrayrulecolor{black}\bottomrule
    \multicolumn{3}{l}{[1] With the term \textit{characterization} we refer to what \citet[p.256]{Briand1996applyGQM} define as ``forming a snapshot of the} \\
    \multicolumn{3}{l}{current state'' of the object of study. In our case, this refers to participants’ shared imaginations of metaverse use} \\
    \multicolumn{3}{l}{cases (Goal 1), or their shared imaginations on future metaverse societies (Goal 2).}
    \end{tabular}
\end{table}

\subsection{Recruitment of Participants}

The question of who participates is central to participatory (speculative) design \cite{Fung_VarietiesParticipation, Farias2022Social}. As \citet{Farias2022Social} report, the recruitment of ``the `right' kind of participants'' \cite[p.2]{Farias2022Social} appears to be a common problem for research projects applying participatory speculative design. To compose a sample of interested citizens, our recruitment method deploys a combination of (a) the most frequently used approach of \textit{self-selection}, enhanced through (b) \textit{targeted recruitment} to also reach (c) \textit{lay stakeholders}\footnote{In the context of participation in governance, Fung \cite[p.68]{Fung_VarietiesParticipation} defines lay stakeholders as ``unpaid citizens who have a deep interest in some public concern and thus are willing to invest substantial time and energy to represent and serve those who have similar interests or perspectives but choose not to participate.''} \cite{Fung_VarietiesParticipation}. We choose self-selection as a strategy based on the principle of universality, which seeks to unite volunteer citizens to collectively contribute towards enhancing the future for everyone and which is a participatory ideal of citizen engagement \cite{mccarthy2023dark}. An open registration process (similar to \cite{mccarthy2023dark}) was utilized to recruit participants from Tokyo and Yokohama, Japan. Information was provided on a workshop webpage and the event registration site peatix.com. Following \citet{mccarthy2023dark}'s recruitment process, who conducted a 16-hour online citizen dialogue, we applied snowball sampling \cite{Goodman1961Snowball} via different channels to specifically target individuals generally interested in technology: we advertised the workshop via word-of-mouth, contacted technology-savvy groups such as Meetup groups or labs in Tokyo and Yokohama via email, and posted workshop registration information on social media channels of the research project, the hosting research lab, the university’s network, and the project’s communication partners. These were \textit{Mutek}, a cultural association dedicated to enhancing digital creativity in music and audio-visual art, and the \textit{Goethe Institute Tokyo}. Both partners and the spread of advertisements were chosen to ensure registrations from different age ranges and individuals beyond the academic sphere and to address people interested in digitalization, a prospective metaverse, and related cultural perspectives. Based on the principle of universality, we welcomed all registered individuals who indicated to be aged above 18 years to participate. We had no intention to perform any screening procedures. We reserved the right to select participants from the registrations only in the event of excessively high registration numbers, however, this situation did not arise.

Participants did not receive a monetary compensation. Incentives to take part were purely of an educational nature, such as learning about the metaverse and how to apply design methods. This non-financial but rather educational incentive or the incentive to jointly explore a topic space to increase agency (within a community) is not uncommon for participatory (speculative) design studies \cite[e.g.,][]{Sharma_2021_Romance, Chopra_2022_Gardening, Lindstrom2020Making, Gerber2018Participatory}. Our workshop advertisement is inspired by these studies. We reflect on the limitations of our recruitment strategy in the limitation section. 

We obtained an ethics approval from our university for conducting this study, including the specifics of our recruitment procedure, the design of the workshop process and the handling of the results. We followed standard practices for ethical research (including informing participants, obtaining consent) while performing the study and analyzing the data. As part of registration, participants gave informed consent to be part of a study. The consent form that participants agreed to when registering to the workshop outlined the study's objectives, i.e., exploring perceptions about the metaverse. It outlined how privacy or other potential risks linked to their involvement in the workshop were minimized. The form articulated that the analysis of the workshop would be conducted on a group basis and that their participation would be kept anonymous and confidential -- no personal information would be recorded during the workshop. Participants were briefed about the potential publication of their workshop contributions, which they gave their permission for.

Initially, after the registration deadline, we counted 35 participant registrations, of which 10 participants actively de-registered after we had to announce switching to a hybrid workshop mode due to an increase in Covid-19 infections in Tokyo/Yokohama. Finally, nine of 25 registered participants did not show up at the workshop (online no-show rate: 42.1\%). 
Also other studies \cite{Oates_2022_interviewmode} experience high no-show rates (\textasciitilde50\%) for online video interviews. In total, an interdisciplinary group of 16 people (10 females, 6 males) participated in the workshop. Participants’ ages varied between 18 and 65 (18-25: 5; 26-35: 5; 36-45: 5; 46-55: 2; 56-65:1). All participants were either native or fluent in Japanese. One participant was originally from China, and the rest from Japan. While three out of 16 participants indicated to be working in the field of design and illustration (see Appendix~\ref{App:Demographics} for all occupations), all participants were treated the same with respect to instructions and roles in the design process. 

\subsection{Workshop Conceptualization and Process}

We applied a PSD approach that was informed by SD and DF practices alike\footnote{With regard to Kozubaev et al.’s notion to utilize Fry’s term design futuring \cite{Fry2009Design} as an umbrella term to refer to various approaches using design to explore futures \cite{Kozubaev2020Expanding}, our approach might as well be called participatory design futuring.}. We organized a two-day PSD workshop in August 2022. The duration amounted to a total of 11 hours (day 1: 5h, day 2: 6h). Initially, with 35 registered participants, we formed eight workshop groups of four to five individuals based on provided demographic information. This means that a group was composed of at least two female and two male participants, that the age range within a group was not greater than three decades, and that occupations varied within the groups, for example, that individuals who indicated to be working for creative industries were spread across all groups. These measures were taken to counteract possible hierarchy effects known from the job market. We reorganized the group setup right before the workshop started, considering the final number of participants to ensure fair distribution of indicated gender and occupation and sensible variation of age within the three online groups (two groups of four participants and one group of three participants). All five people participating on-site formed one group. Due to an increase in Covid-19 infections in Tokyo/Yokohama, only one participant group was pre-selected (first-come-first-serve taking demographics into account) and allowed to participate in person to adhere to safety measures. It was emphasized that everyone participated as an everyday expert and that no participant had knowledge advantages about the future. Live communication across groups online and on-site was facilitated via Zoom. To support constant communication and easy accessibility, the Zoom call remained active throughout the entire duration of each of the two workshop days, including during the lunch break, allowing all participants to interact with each other and readily contact facilitators whenever needed. The entire workshop was held bilingual (English, Japanese), including oral presentations and written workshop material.

To guide the workshop, we designed a digital toolkit (see Appendix~\ref{App:WSToolKIT}) transporting the framework that allowed participant groups ``to make artefacts about or for the future. [It gave] […] non-designers a means with which to participate as codesigners in the design process'' \cite[p.9]{Sanders2014Probes}. Specifically, the toolkit guided groups to create magazine articles from the future. This format was chosen to give groups visual and textual freedom to materialize the created narrative in a manner that they felt most comfortable with. It provided a way to concretize the otherwise speculative conversations the participants were having, similar to the approach of \citet{Elsden2017On}. The decision to choose this narrative approach is grounded on prior research \cite{LIVELEY2021102663} illustrating the importance of narratives as foundation of futures thinking and their potential to enhance critical reflexivity. The toolkit guided groups to create artifacts “designed to provoke or elicit response” \cite[p.9]{Sanders2014Probes} from an external audience in follow-up discourses. The process of creating the magazine article (see Figure~\ref{fig:instructionsPrototype} in Appendix~\ref{App:prototyping} for detailed instructions) can be referred to as fictions as participatory constructions, whereas the use of the created magazine article in follow-up discourses can be considered as fictions as probes \cite{Muller2017Exploring}; drawing on \cite{Sanders2011Experiencing}. 

The workshop framework concept is inspired by participatory/co-(speculative) design and design fiction processes (e.g., \cite{Epp2022Reinventing, Harrington2021Eliciting, Light2021Collaborative, Sanders2014Probes, Ventä-Olkkonen2021Nowhere}). It has also been complemented through two authors’ practical knowledge of service design methods. The concept builds on what \cite{Brandt2006Designing, Jungk1996Future} describe as \textit{future workshops}, taking into account the design choices framework \cite{Lee2018Design} for co-creation processes. The rationale behind the chosen structure and methods (see Table~\ref{tab:workshop_outline}) was to guide participants’ thoughts without pre-framing, giving them the opportunity to apply their individual perceptions to questions at hand. This was inspired by Light’s \cite{Light2021Collaborative} concept of \textit{seeding instead of leading}. The toolkit was deployed using the collaborative digital platform \textit{Mural}, where groups documented their process. 

As proposed by \cite{Sanders2014Probes}, the process of using the toolkit was facilitated, here by five authors\footnote{While four researchers conceptualized the workshop and were tasked with forming the groups, another researcher joined as of the preparation phase of the workshop, and participated in facilitating the workshop and analyzing the results.}, who did not directly participate but observed groups’ processes. The authors introduced each new step of the toolkit and were continuously available in case of questions. Table \ref{tab:workshop_outline} presents the four workshop phases, namely, understanding, speculating, creating, prototyping, as well as the corresponding tasks, outputs, methods and sources of methodical inspiration. The first phase included creating a group profile, performing desk research on the metaverse, and jointly formulating a definition. The desk research to create a joint knowledge base was adapted for the use of non-designers from \cite{Stickdorn2018ServiceDesign}. Groups’ joint definitions of the metaverse served as summary of the activity. Researching about metaverse technology and related socio-technical trends was also partly inspired by \cite{Epp2022Reinventing}’s scanning activity to kick-off a speculation workshop. This task of jointly performing research and defining the metaverse also served to ensure construct validity \cite{Wohlin2012GQM} within the groups, i.e., that group members had a similar idea of the concept of the metaverse. The second phase consisted of a guided imagination and an exploration exercise, inspired by \cite{Soden_facct_2022}. As we were also interested in groups' different understandings, we did not strive for construct validity across groups. Participants were guided in their imagination of a future society through a read-aloud narration. Speculations were mapped on the STEEP framework \cite{SchwartzSTEEPLE1991} as in \cite{Epp2022Reinventing} – i.e., how can the speculations be categorized in relation to the categories Sociological, Technological, Economic, Environmental and Political (S.T.E.E.P.). Groups selected one dominant idea and performed a consequence mapping to form a scenario world. These activities were a continuation of \cite{Epp2022Reinventing}’s scanning activity and the start into what is similar to their ripple creation process. In the third phase, the groups brainstormed products or services that were either part of the previously imagined future scenario or played a central role in it. Mapping the ideas onto an actantial model \cite{Greimas1984Structural}, originally a method for narrative analysis, helped to sharpen the idea by identifying all relevant actors in the context of the service/product concept. The use of this framework for a creation process is an adaption from previous work, where it was used to analyze speculative design artifacts \cite{Hohendanner2021Designing_not_anonym, Hohendanner2023ReflectiveSpace}. The adapted use of the model can also be seen as a simplified version of a service design method called value network maps \cite{Stickdorn2018ServiceDesign}. The exercise resulted in the joint formulation of a service/product concept through a value proposition, another method from the service design business model canvas \cite{Stickdorn2018ServiceDesign}. In the final phase, groups created magazine articles (see Figures~\ref{App:FigDisaster} to \ref{App:FigXRFood} in Appendix~\ref{App:MagazineExerpts}), which were presented and discussed in an open forum. This phase was inspired by prior works' \cite{Sharma2021From, Prost2015From} scenario writing exercises and adapted with elements from one additional service design method called service advertisement \cite{Stickdorn2018ServiceDesign}: As participants crafted their narratives in the format of magazine articles, they were required to consider all the nuances associated with engaging an imagined audience, akin to the process of creating mock advertisements for a hypothetical readership.

\begin{table}[hbt!]
    \caption{Workshop phases, tasks, methods and output}
    \label{tab:workshop_outline}
    \small
    \begin{tabular}{p{0.07\linewidth} p{0.3\linewidth} p{0.15\linewidth} p{0.2\linewidth} p{0.16\linewidth}}
    \toprule
        Phase & Guiding Question$/$ Task & Methods & Output & Informed by\\ 
    \midrule
        1) Understanding & What is the metaverse? & Group profile, desk research, key sentence & Written definition of how the group understands the concept of metaverse & Prep research \cite{Stickdorn2018ServiceDesign} \& scanning exercise \cite{Epp2022Reinventing} \\ 
    \arrayrulecolor{lightgray}\hline       
        2) Speculating & What if the metaverse was an integral part of a future society? & Guided imagination, group discussion, STEEP, consequence mapping & Visual mapping of various facets and societal consequences of prospective metaverse technology & Narrated imagination \cite{Soden_facct_2022}, scanning exercise \& creation of future ripples \cite{Epp2022Reinventing} \\
    \hline     
        3) Creating & Which metaverse-related product/ service could play a central role in the speculated future? & Brainstorming, actantial model, value proposition & Written concept of a fictional metaverse service or product & Actantial model \cite{Greimas1984Structural, Hohendanner2021Designing_not_anonym, Hohendanner2023ReflectiveSpace}; value proposition (business model canvas) \cite{Stickdorn2018ServiceDesign} \\ 
    \hline     
        4) Prototyping & How could the speculated product/ service \& related scenario be depicted in a magazine article in the future? & Prototyping, presentation and open forum & Digital magazine article & Scenario writing \cite{Sharma2021From, Prost2015From} \& ad-prototyping \cite{Stickdorn2018ServiceDesign} \\ 
    \arrayrulecolor{black}\bottomrule
    \end{tabular}
\end{table}

This approach allowed a two-way sensemaking process among participants: On the one hand, the use of PSD allowed to jointly investigate the capacities and limits of a technology (= sensemaking of technology). This process was triggered by imagining the metaverse to be an integral part of society (second phase), hence, to hold a central socio-technical position. This results in an attribution of solution competence for central societal problem areas. On the other hand, participants identified social problem areas of the future that they perceived as central in their cultural and local context (= sensemaking of the future through technology). A discourse on the question of how we want to live together in the future emerged.

\subsection{Data Collection and Analysis Approach}

We deliberately decided to neither record Zoom calls nor on-site group discussions. While recording interviews or focus groups is often taken for granted \cite{tuckett2005recording}, 
study experiences show that recording conversations can intimidate study participants and influence what is said \cite{Nordstrom_2015_Recording, Caronia_2015_Recording}. Not recording can contribute to a more open and natural conversation \cite{Krauss_2003_MediaResearchInJP, Rutakumwa_2020_Interviews}. More important than recording the precise wording as the foundation for data collection, is enabling an environment in which participants feel at ease expressing their thoughts on a specific subject. To increase participants' levels of comfort and to ensure they were supported throughout the workshop, Japanese-speaking facilitators regularly visited the groups after asking for permission to interrupt. They clarified questions on the tasks and the mode of documenting their discussions in the workshop material if participants expressed uncertainties. Besides facilitator-initiated visits, participants could reach out to facilitators at any time.

The collected data includes the workshop material, where participants documented their progress, and the created design artifacts, i.e., the magazine articles including the visualizations (three groups used free generative AI tools to create their visualizations, one group used freely available stock images). The main language of the material is Japanese. Some groups documented tasks partially in English. All data was used for the analysis. Five researchers, three born and raised in Japan and two born and raised in Germany, jointly performed the qualitative analysis during three analysis workshops to identify emerging themes. Each researcher brought a distinct set of methodological experiences to the analysis process: Three researchers have a strong background in qualitative coding with international researcher teams, two in analyzing design artifacts, and three in analyzing speculative or science-fictional workshop results. The narratives were translated into English by one Japanese author who also works as a professional interpreter. Table~\ref{tab:summary_narratives} briefly summarizes the four fictional narratives. Appendix~\ref{App:MagazineExerpts} provides more detailed descriptions of the four fictional narratives summarized by the five researchers during the analysis sessions with visual impressions from the created magazine articles.

\begin{table}[hbt]
    \caption{Summary of fictional narratives created during the workshop}
    \label{tab:summary_narratives}
    \small
    \begin{tabular}{p{0.08\linewidth} p{0.87\linewidth}}
        Group & Summary of Fictional Narrative \\
    \midrule
        Disaster Prevention &  The article reviews a trial evacuation training in the metaverse, called ``disaster simulation''. This service simulates natural disasters, allowing users to practice evacuations in virtual replicas of their real-world environments, with feedback for improvement. The author praises its potential for real-life disaster preparedness but raises concerns about VR-induced trauma and desensitization to violence. The service also aims to analyze user behavior during disasters to aid governmental emergency preparedness. \\ 
    \arrayrulecolor{lightgray}\hline       
        MOTHER & The article reviews MOTHER, a new avatar generation service, set in a future where AI-driven avatars in the metaverse act autonomously, representing their users. Central to the story is the increasing autonomy of these new avatars, raising questions about granting them citizenship rights. Perspectives vary: users are divided on civil rights for avatars, an engineer reflects on past service issues, a real-world advocate calls for control over avatars, and a woman worries about her family's reliance on avatars. The article suggests these debates are set to intensify. \\
    \hline     
        U12-Topia & The article addresses whether ``collective personalities'' – AI representations of metaverse community opinions – should have political rights such as being elected. The metaverse is described as a platform where users form AI-driven collective personalities through discourse within separate communities, e.g., the community of schoolchildren ``U12-Topia''. In a future with a politically dominant elderly population, the article debates the inclusion of younger generations in politics through these AI entities. Perspectives range from a developer advocating for societal creative agency, a schoolchild emphasizing youth inclusion in politics, to a legal expert stressing the need for discussing the rights of collective personalities. \\
    \hline     
        XR-Food & The ``XRFood'' magazine issue showcases the impact of XR (extended reality) food technology in a world recovering from a food crisis. Featuring articles and ads by various characters, including researchers and an XR restaurant owner, the issue focuses on a metaverse restaurant that offers unique taste experiences globally. While XR technology solved food scarcity and health issues, it led to the loss of diverse food cultures. The magazine highlights XR food's benefits in entertainment and medicine, allowing people with dietary restrictions to enjoy any food and offering anonymous dining experiences in the metaverse to help mend post-crisis relationships.  \\ 
    \arrayrulecolor{black}\bottomrule
    \end{tabular}
\end{table}

First, we performed the analysis at the descriptive level in analysis workshop 1, in which five researchers took part. Consulting all data collected (groups' narratives and workshop documentations), we analyzed the metaverse product or service at the center of the narrative and how it affects the depicted future scenario. A comparative mapping of the narratives according to the actantial model \cite{Greimas1984Structural} built the foundation for the analysis and revealed similarities and differences in the narratives. The actantial model allows mapping who is the main subject of a story (subject), what goal the subject wants to achieve (object), on whose order is the subject acting (sender), who is opposing the action (opponent), who is aiding the subject in the action (helper), and who is benefiting from the action (receiver). To exemplify, Table~\ref{tab:actantial_model} in the Appendix~\ref{App:ActantialModel} presents the actantial models for two narratives and provides more information on the analysis procedure. Actantial models as the main outcome of the descriptive phase informed the follow-up analysis to answer the questions characterizing our measurement goals (see Q1.1-Q2.2 in Table \ref{tab:GQM}). 

\begin{figure}[h]
  \centering
  \includegraphics[width=\linewidth]{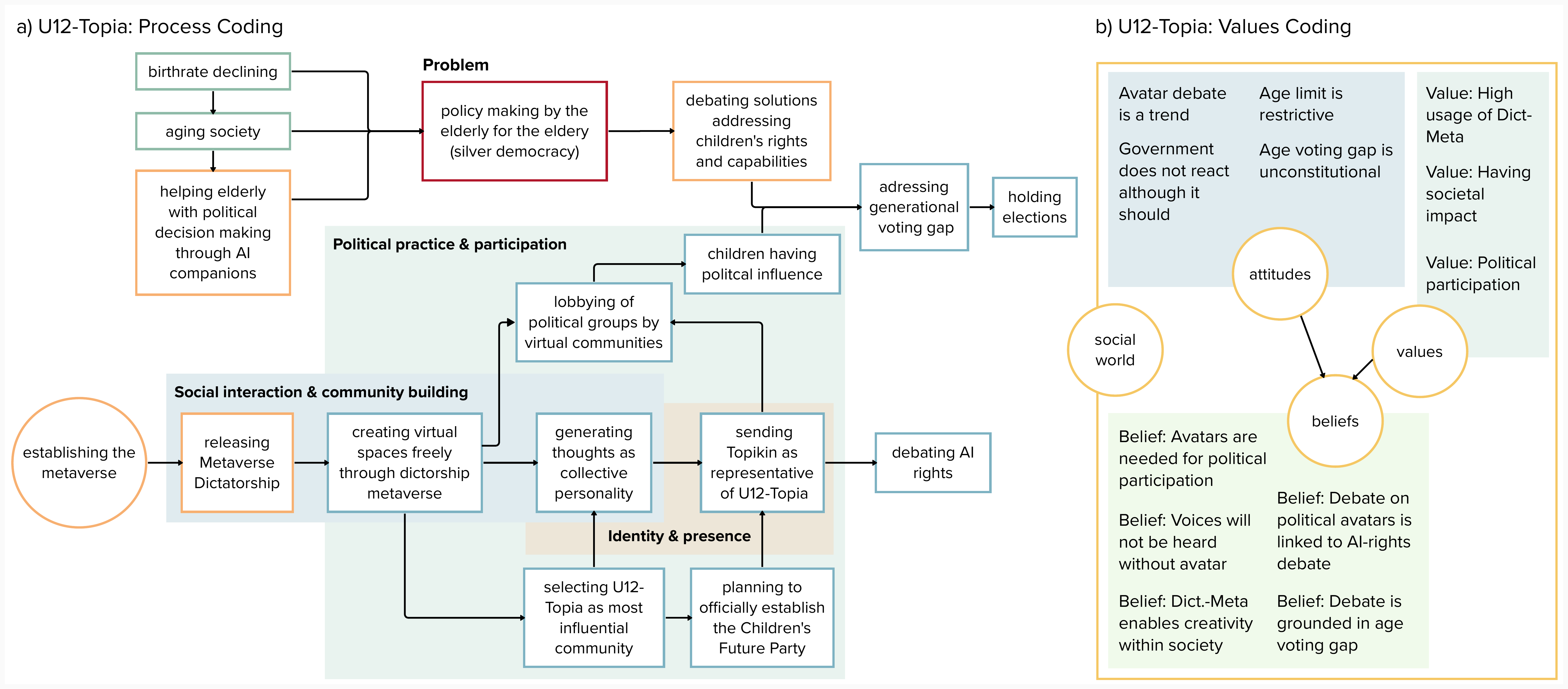}
  \caption{Example of process coding and values coding for the workshop group “U12-Topia”; Source: Own illustration.}
  \Description{Example of process coding and values coding for the workshop group “U12-Topia”; Source: Own illustration}
  \label{fig:process_analysis}
\end{figure}

Second, we entered the deriving level to identify superordinate themes within the narrative and the scenario during two further analysis workshops. To do so, the narratives were analyzed through two rounds of coding (one round before each of the two workshops), each followed by joint discussions of identified codes and refinement as well as alignment during the two workshops (see Figure~\ref{fig:process_analysis}). Process coding \cite{Saldaña2013coding} allowed us to identify how metaverse development was depicted in relation to prospective societal problem spaces (Q1.1) and potential problem-solving capacities (Q1.2). Process coding was also applied to identify anticipated transformation within speculated metaverse societies (Q2.1). Process coding can be utilized to mark action observed in the data using gerunds: Both, observable activities as well as conceptual activities can be process coded, and be ordered sequentially or visualized graphically \cite{Saldaña2013coding}. Values coding \cite{Saldaña2013coding} allowed us to map values, attitudes and beliefs depicted with regard to metaverse technology and prospective societies (Q2.1 \& Q2.2): Values reflect the importance that is attributed to an object of reference (idea, thing, person, self); attitudes describe the way we think about these objects of reference; beliefs can serve as ``rules for action'' \cite[p.28]{Stern2011GTEssentials} and are part of a system composed of attitudes and values, as well as opinions, experiences, knowledge or other interpretative subjective perceptions \cite{Saldaña2013coding}.

Identified codes were organized into superordinate themes based on patterns observable in the data: Themes were identified by comparing the actantial models of the four narratives, looking for similarities in actants and their relations. Through process coding, we identified developments of the actants and their relations with regards to problem spaces and respective solution competences of metaverse technology (see Section~\ref{Sec:sensemaking_of_metaverse}), making them comparable between narratives. Process coding also revealed societal transformation processes inherent in the narratives and made them comparable across the narratives. This was complemented by values coding to uncover arising societal value discourses (see Section~\ref{Sec:sensemaking_of_future}). To formalize the findings, we adopt a quasi-statistical approach \cite{Robson_2002_Research} to report in how many groups we find a superordinate theme.

\section{ANALYSIS OF UNDERLYING THEMES}
In this section, we present identified themes resulting from the researchers' joint analysis sessions. 

\subsection{Sensemaking of Metaverse Technology} \label{Sec:sensemaking_of_metaverse}

We identified four themes based on respective problem areas in which participants attribute problem-solving capabilities to a metaverse of a fictitious future society (see Table \ref{tab:sensemaking_metaverse}). 

\begin{table}[hbt]
    \caption{Sensemaking of the metaverse (attribution of solution competence)}
    \label{tab:sensemaking_metaverse}
    \small
    \begin{tabular}{p{0.15\linewidth} p{0.3\linewidth} p{0.48\linewidth} }
    \toprule
        Superordinate Theme & Problem & Metaverse as solution \\
    \midrule
        Social interaction and community building & Limitations of physical environments for communication and social interaction & - Metaverse as open digital space for expression and social interaction \\ 
    \arrayrulecolor{lightgray}\hline     
        Identity and \newline presence & Societal expectations and overwhelming number of tasks and interactions that come along with the simultaneous presence in the physical and digital world & - Metaverse as space to appear with self-selected identities \newline - Avatar as representation of digital self \newline - Semi-autonomous avatars help take on the plethora of interactions and tasks of individuals \\
    \hline     
        Political practice \& participation & Imbalance of citizen representation in political practice (silver democracy) & - Platform for discourse and opinion building processes \newline - AI-driven collective personalities as output medium connecting to institutionalized political sphere \\
    \hline     
        Virtual training space and extension of senses & Limitations of physical environments and human senses to prepare and adapt to crises &- Simulation space for real-world environmental disasters \newline - Stimulation of gustatory senses to adapt to food supply and health concerns \\ 
    \arrayrulecolor{black}\bottomrule
    \end{tabular}
\end{table}

\textit{Social interaction and community building.} A common theme of all four narratives are forms of community building and moderation as well as increased possibilities of social interaction in light of real-world challenges or limitations of physical environments. In U12-Topia, the metaverse is depicted as a space for community building, where peers can come together and design their own virtual environments. Here, like-minded people can overcome the restrictions of real world environments, e.g., school children can discuss, train and carry out political actions. In the narrative MOTHER, social relationships and associated communities are forged across physical and digital spaces. Autonomous avatars in the metaverse acting on behalf of their owners allow to tie new social bonds with other users and their avatars alike, expanding social relations autonomously according to anticipated user needs. The XR-Food project proposes shared experiences of technologically enhanced eating in the metaverse, building a global food community to overcome social tensions after global food crises. Finally, the Disaster Prevention project offers joint training experiences as social activity in the metaverse with others to increase preparedness of real-world communities for natural disaster events.

\textit{Identity and presence.} Two narratives center around questions about the influence on human identity that come along with the possibilities of virtual representation of the human self or one's opinions and beliefs. In the narrative MOTHER, the possibility to create AI-driven semi-autonomous avatars as representation in the metaverse is described. The avatars serve as a solution to the issues arising from the overwhelming number of tasks and interactions that come along with humans' simultaneous presence in the physical and digital world. Avatars, rather than being representations of an individual, become an extension of the human self, as they can act on their own, capable of meeting the needs of communal spaces where digital and physical realms converge. For example, avatars can perform parental duties by looking after their owner's child who is spending time in the metaverse. The project raises the question whether outsourcing of the self using AI technology is a viable and desirable solution. Furthermore, both narratives U12-Topia and MOTHER highlight the possibility to appear in the metaverse with self-selected identities\. U12-Topia takes this principle one step further, as the narrative describes the possibility of whole virtual communities being represented by a single AI-driven avatar, channeling all discourses, beliefs and values present in the virtual community into one entity. The personalized representation of the digital self parallels discourses in VTuber and VR communities. This enables users to behave and act in ways of their choosing in virtual spaces. Thus, the avatars ``provide access points in the creation of identity and social life'' \cite[p.40]{Taylor2004Modern}.

\textit{Political participation and practice.} In the narrative U12-Topia, the metaverse becomes a political discourse space for younger generations. It allows for political opinion formation processes through the generation of AI-driven collective personalities connecting to institutionalized political processes. The narrative describes the lack of participatory approaches in society and politics, (digital) isolation, and an aging society that largely determines political processes as a source for these developments. By providing U12-Topia as a technical means for a participatory place where any user can find or create a representation of an ideal society or community, U12-Topia aims at confronting the problem of Japan’s silver democracy \cite{McClean2020Silver}. This term describes a democracy in which the elderly make up a majority of the electorate, and are also over-represented within political institutions. Thus, they can exert greater influence on the political sphere \cite{McClean2022Generational, Okamoto2020Silver}, but give less consideration to the needs of the younger generations leading to a generational imbalance \cite{Seo2017Democracy}. 

\textit{Virtual training space and extension of senses.} Three narratives describe forms of manipulation or an extension of human senses and discuss their potential consequences. In the narrative MOTHER, autonomous avatars serve as extension of the human self, as they can make experiences in the virtual realm autonomously that can later be revisited by their human users. Also the narrative XR-Food describes the extension of senses as the consumption of food is enhanced by the manipulation of human senses through XR-Food technology. The Disaster Prevention narrative describes the metaverse as a space suited to simulate dangerous natural phenomena in a safe way. The mutual theme of these narratives is the idea of the metaverse being a space for making virtual experiences as a way of testing real-world challenges (training) without having to deal with potential negative consequences or as a way to extend human senses beyond physical limitations.

\subsection{Sensemaking of the Future Through Metaverse Technology} \label{Sec:sensemaking_of_future}

Workshop results discuss the following societal issues and value implications of a fictitious future metaverse society across five identified themes (also summarized in Table \ref{tab:sensemaking_future}). Participants identified these as important to be negotiated in the course of the development of a prospective metaverse and future societies.

\begin{table}[h]
    \caption{Sensemaking of the future through the metaverse}
    \label{tab:sensemaking_future}
    \small
    \begin{tabular}{p{0.13\linewidth} p{0.32\linewidth} p{0.48\linewidth}}
    \toprule
        Superordinate Theme & Representation of Anticipated Transformation within Speculated Metaverse Society & Societal Questions and Issues, and Value Discussion \\
    \midrule
        Human-AI\newline relations & Existence of AI entities that take on human tasks, fulfill societal roles and build social relations & - AI rights \newline - Legal and moral responsibility \newline - Identity and autonomy \newline - Alienation \newline - Dependence on technology \newline - Experience and task overload \\ 
    \arrayrulecolor{lightgray}\hline    
        Technological\newline solutionism & Metaverse technology as universal problem solver & - Reliance on technology \newline - Human control over new emerging technologies \newline - Social divide between technology users/adopters and skeptics \\ 
    \hline    
        Politics and\newline technology & Wish of political participation through smart opinion representatives & -	Political representation \newline - Need for political system to adapt to the development of new technologies \\
    \hline    
        Globalization and local cultures & Universalization of (food) cultures through global communication and shared practices & - Effect of ongoing globalization on local practices \\
    \hline
        Immersive\newline technology & Simulation of hyper-realistic scenarios and stimulation of senses & - Loss of sense for and connection to reality \newline - Escapism from reality \newline - Dulling of feelings \newline - VR traumas and post-traumatic experiences \\
    \arrayrulecolor{black}\bottomrule
    \end{tabular}
\end{table}

\textit{Human-AI relations.} As a central transformation within the societies in the narratives around U12-Topia and the service MOTHER, we observe the relations between humans and (semi-)autonomous AI entities. Given that AI entities fulfill societal roles and build social relations, the narratives raise questions on identity and autonomy, both of human and AI entities. Further, questions of human dependence on technology and the alienation from other humans arise. The theme substantiates the question whether AI entities should be given rights and who is responsible in case of problems. The social pressure to be active in both virtual and physical spaces might lead to overstimulation and task overload, increasing the need for AI support.

\textit{Technical solutionism.} In the XR-Food magazine, we observe the promise of technology being the cure to numerous problems, ranging from food shortages, people's lack of essential nutrients and even world peace. In this context, high value is attributed to the technological solution. Yet, the workshop group named the restaurant in their narrative DYSTOPIA, hinting at negative aspects of this development. U12-Topia and the service MOTHER question how far-reaching human control over new emerging technologies is. The future scenarios present a reliance of humans on technologies. The projects challenge the belief that technology is the solution to (social) problems. Follow-up problems are identified: A societal divide is anticipated between technology adopters and skeptics.

\textit{Politics and technology.} The influence of technology on future political practices represents a key topic in the U12-Topia narrative. There is a social change in society that is reflected in the desire for youth political participation, which is presented as an important value for the youth of the fictional society. This is grounded in the government's behavior of not appropriately reacting to the effect of the aging society on the realities of political participation. Questions of future political participation supported by digital innovations such as AI-enhanced opinion-forming processes leading to, for instance, AI-supported representative democracies, and the integration of younger generations in political processes as reaction to aging societies are negotiated. In addition, the phenomenon of echo chambers in digital opinion-forming spaces is addressed.

\textit{Globalization and local cultures.} In the XR-Food project, universalization of (food) cultures illustrating the replacement of local food cultures by a global one represents a transformation affecting the depicted society. This can be seen as an expression of how ongoing globalization might affect local practices, and of the high value that is attributed to the preservation of local (food) cultures.

\textit{Immersive technologies.} All projects highlight aspects of the effect of immersive technologies as a core transformational element of future societies on human behavior. The narratives weight the benefits and disadvantages of using immersive technologies that allow the simulation of hyper-realistic scenarios and stimulation of senses. On the one hand, users can build stress tolerance and train or increase their capabilities. On the other hand, negative effects such as the loss of the sense for reality, dulling of feelings or VR traumas and post-traumatic experiences after experiencing hyper-realistic disasters and violence in virtual spaces can emerge.

\section{Discussion}

The present study used PSD to explore what issues, questions and values have to be discussed for socially viable and beneficial ways to conceptualize, develop, and adopt a prospective metaverse. In the following, we discuss our findings by, first, contextualizing the results with regard to metaverse for social good application areas. Second, we raise questions to metaverse designers, media, and politicians, and third reflect on our methodological approach.

\subsection{Discussion of Identified Themes} \label{Sec:dicussion_themes}

On the one hand, some of the topics from workshop narratives can be assigned to those identified in general discourse. On the other hand, some topics refer to local parameters that are specific to Japan. 

With regards to the impact of the metaverse for the social good application areas diversity, accessibility, equality, and humanity \cite{Duan2021Metaverse}, we observe that participants discuss these topics from a more critical perspective. For accessibility and diversity, Duan et al. \cite{Duan2021Metaverse} highlight the metaverse's potential of providing spaces of interaction and expression without physical limitations to a broad variety of users. The potential dangers of these circumstances, e.g., the permanent need for virtual presence or the resulting number of social interactions and information that can no longer be handled by individual users, are discussed by MOTHER. Likewise, U12-Topia discusses accessibility in the context of political processes, aiming for a diversified political sphere, but only under conditions of full dependence on technological enhancement in form of AI.

Duan et al. \cite{Duan2021Metaverse} further identify equality as a result of questioning social constructs through the use of avatars to be one positive impact of a metaverse for social good. Supporting this claim, research shows that VR platforms provide new opportunities to approach various gender identities \cite{Freeman2022(Re)discovering}, and can support LGBTQ community members in meeting their needs to the extent of providing a safe space \cite{Acena2021“In}. While not referring to gender identity specifically, MOTHER and U12-Topia allow users to self-select their virtual identity. This usage of avatars reminds of the hyperpersonal model \cite{Walther2021Language}. Applied to VTubers, the model suggests that senders using virtual avatars hold advantages when interacting with receivers compared to appearing in person \cite{Lu2021More}. This advantage results from the possibility to modify one’s self-representation by selectively choosing how to present oneself, i.e., which cues to send \cite{Lu2021More}. However, the possibility depicted by the narrative MOTHER to continuously extend one’s identity through the experiences of the avatar endangers the human user also to lose control over identity formation.

The narrative XR-Food simultaneously contrasts and aligns with existing research: While earlier studies like \cite{Duan2021Metaverse} suggest the metaverse's potential in preserving cultural achievements and enhancing accessibility, XR-Food presents a different view, illustrating the potential replacement of local cultures with a globalized one. However, this narrative aligns with ongoing research themes about the influence of virtual environments on food experiences and perception, as explored by \cite{Taguchi2023VRTaste} and \cite{Nakano2022OverlayEating}. It also resonates with discussions on how the metaverse could transform our relationship with food, as indicated by \cite{Covaci2023DigitalPie}. This duality in XR-Food's narrative reflects the complex and multifaceted impact of the metaverse on cultural practices and experiences.

The narratives U12-Topia and MOTHER present various perspectives on rights for AI agents; however, they are ultimately also highlighting anxiety about the development of autonomous AI entities. Robot and AI rights are globally discussed topics and find both supporters \cite{Koops2010Bridging,Turner2019Robot} and critics \cite{Bryson2010Robots,Bryson2017Of,Solaiman2017Legal}. Research finds that US citizens are, at first, not in favor of AI and robot rights, however, they also dislike punishment and cruel treatment \cite{Lima2020Collecting}. Closely related to this is the public perception of AI, which is subject to common narratives about AI \cite{Bareis2022Talking,Cave2018Portrayals,Cave2019Hopes}. Research finds cross-cultural differences in the perception of AI between citizens from Japan and the US, or Japan and Germany~\cite{Ikkatai2022Segmentation,Nadeem2020Science}.

Themes with specific cultural references evolve around the silver democracy, disaster prevention and the preservation of food culture. With regards to the silver democracy and problems of political participation of the youth in Japan, researchers discussed electoral reform \cite{Okamoto2020Silver}. One example is the introduction of the so-called Demeny \cite{Demeny1986Pronatalist} electoral system allowing parents to vote for their minors who obtain passive voting rights \cite{Aoki2009Is,Okamoto2020Silver,Vaithianathan2013Support}. U12-Topia develops a form of AI-driven Demeny in which electoral rights are granted to the AI entities (collective personalities, e.g., Topikin) that channel the discourses of schoolchildren communities. Interestingly, the name of the collective personality Topikin bears strong similarities to Hikakin, currently one of Japan’s most influential YouTubers \cite{Abidin2021Influencers}. In the wake of measures to contain the COVID-19 pandemic, Hikakin conducted a much-noticed interview with Tokyo’s Mayor Koike. This led to debates about influencers being part of political communication strategies, especially engaging younger audiences~\cite{Osaki2020Lets,Suzuki2020Tokyo}.

The fictional Disaster Prevention project reflects recent ambitions making use of VR to observe people’s reactions, behavior, and stress levels during emergencies in order to increase safety \cite{Gamberini2021Designing}. This is especially valuable in geographic regions where natural disasters occur frequently, such as in Japan “where people are required to be alert against disasters” \cite[p.301]{Ooi2019Virtual}. The need for disaster education in school arises \cite{Ooi2019Virtual}, and teaching material is provided by the Ministry of Land, Infrastructure and Transport \cite{MLITJapan2022Introducing}. The current range of natural disaster VR simulation training systems can be differentiated into tools for mitigation and preparedness, e.g., training for (fire) disaster evacuation \cite{Ooi2019Virtual,Ren2006Application,Xi2014Simulating}, and response and recovery \cite{Li2022Review}. The workshop group’s idea is in line with the perception that new computer technology allows for “a new way of thinking for disaster emergency management research” \cite[p.1875]{Li2022Review}. 

\subsection{Implications for Metaverse and AI Development}

From the issues identified, questions for metaverse designers and engineers, as well as for media and policymakers arise, which should be considered when striving for socially sustainable developments and resilient future societies: While even today people struggle with problems emerging from current information societies, we should consider what effects might result from the emergence of more social VR applications. How should we deal with the highly increased demand for attention and interactions that the simultaneous presence of individuals in digital and physical spaces requires? How should we handle the potential effects on human perception such as losing touch with reality, and behavior like digital escapism that comes along with increased intersections of physical and virtual spaces and innovations in XR technology? Let us imagine the widespread use of AI entities. How should we deal with systems, acting on behalf of users or autonomously, and potentially not being distinguishable from human users? Today, we see politicians entering virtual spaces such as Instagram or TikTok to reach their electorate. How will political (democratic) systems have to adapt to virtual spaces with regard to new forms of communication, discursive spaces, and opinion formation processes, in order to reach and involve citizens or to respond to them adequately? Likewise, how can we make the benefits of global communication and exchange accessible to everyone without threatening local practices and their preservation in light of technological progress? Finally, from a macro perspective, how can we ensure not to strengthen the logics of technological solutionism despite the highly needed societal transformation processes with regard to global crises such as global warming? 

Our research underscores the importance for a multi-faceted approach to metaverse development that is grounded in inclusivity, ethical considerations, and cultural sensitivity. CSCW researchers, developers and designers of the next generation of metaverses should prioritize creating environments that foster inclusive and equitable interactions, considering factors such as accessibility, representation, and the preservation of local cultures amidst globalized virtual spaces. 
Given that a lack of studies including non-Western perspectives on technology development has been identified by previous research \cite{linxen_2021_WEIRD_CHI, Septiandri_2023_WEIRD_FAccT}, integrating a broader range of perspectives is vital for socially sustainable metaverse development. By incorporating these diverse perspective, future research can pave the way for metaverses that have been critically reflected upon and not simply propose presumed solutions to current societal challenges. Instead, for metaverses that have been critically reflected upon, developing teams should be able to proactively anticipate and mitigate potential risks associated with the merging of physical and virtual realities.

Our approach can contribute to these endeavours. The creation of narratives enables reflection and expression of laypeople's sensemaking processes towards a prospective metaverse. The created narratives can assist individuals in identifying and negotiating specific topics or aspects they're skeptical about and in reflecting upon potential solutions or acceptable alternatives. In this, our approach can help to go beyond mere (public) feelings of scepticism or enthusiasm towards a prospective metaverse. The narratives and emerging themes may point to areas where technical solutions may be required, as they might "express the situation of an alternate present or possible future as an issue: a situation that is contestable.” \cite[p.119]{DiSalvo2012Provocative}. If so -- in the words of DiSalvo, “to be truly provocative is to rouse to action" \cite[p.119]{DiSalvo2012Provocative} -- developers are addressed with a call to action to propose technical solutions to the discovered issue spaces.

Referring to our two-fold contributions, we propose a first of two possible approaches (see second in \ref{sec:Dis_PSD}) for how the results of this study can be utilized: 
(1) Building on the emerging narratives and themes from this paper's research results: (1.1) Researchers can reflect how their research ties in with citizens' broader imaginations and whether their work could enhance positive or rather negative aspects of proposed narratives. This can also uncover where to prioritize design efforts to respond to users' beliefs, perceptions and needs. (1.2) Narratives emerging from our workshop and its analysis can be used to explore where to intensify research on themes that have been explored in previous work (e.g., implications of social VR representation/avatars, human-AI relations in virtual environments). Also, newly emerging themes like political participation through metaverse can direct new research endeavors.

\subsection{Reflecting the Metaverse as Sociotechnical System through Participatory Speculative Design} \label{sec:Dis_PSD}

As our analysis shows, the applied method was effective in encouraging the participants to jointly discuss the social and ethical implications of a prospective metaverse society in depth. Participants were able to explore the potential benefits and risks of the metaverse and associated AI technologies, and to consider potential impacts and related discourses such as human-AI relations, technological solutionism, globalization, and local cultures. The addressed problem areas and raised societal questions also highlight that purported benefits illustrated by previous research need to be critically reflected.

Furthermore, the applied design process enabled participants to contribute with their personal and local background to the imaginary. This is exemplified by themes of political reform in light of Japan’s silver democracy or natural disaster prevention. By uncovering these culturally and locally individual factors and by highlighting particular themes, discussions on how new kinds of worlds could be made possible are enabled. Therefore, the presented approach can serve as an example of prosocial modes of engagement and can inform the CSCW and HCI community, and the general public alike. In this sense, the presented approach might be valuable to Stilgoe et al.’s \cite{Stilgoe2013Developing} framework for responsible innovation. It is applicable for the dimensions of anticipation (as a method to build scenarios and explore visions), reflexivity (as a method for challenging the status quo and taken-for-granted assumptions), and inclusion (as a method that does not require participants to have technical knowledge) \cite{Stilgoe2013Developing}.

Hence, referring to our two-fold contributions, we propose a second possible approach for how the results of this study can be utilized: 
(2) Building on our methodological contribution, our framework (toolkit available in Appendix~\ref{App:WSToolKIT}) can be applied by other researchers. (2.1) Here, one aim should be to conduct more PSD workshops with citizens globally to explore multiple perspectives and identify diverse local culturally-specific themes to further broaden discourses instead of focusing on single aspects of technology. (2.2) Researchers developing new technological applications can apply the framework to explore the implications of prospective societal impact of their technology by involving other researchers or even citizens. Instead of placing metaverse technology in the center of the speculation process, researchers can adapt the central question and pose, "What if <researchers' technological application> was an integral part of a future society?".

\section{Limitations and Future Research}

Our research is not without limitations. Given our participant recruitment strategy, our sample is not representative of the Japanese population. We intentionally chose an online-channel-focused advertisement strategy to reach out to people likely interested in and more knowledgeable about technology. We acknowledge that in this way we exclude people less aware of current technological developments. Additionally, the lack of financial compensation skews the sample towards people who can afford to join an 11-hour workshop on two weekend days. 

The analysis is subject to interpretation which builds upon the researchers’ individual backgrounds \cite{Schlesinger2017Intersectional}. We encourage the reader to join the reflection of the narratives and interpret them as well. The speculations presented in this paper are not exhaustive, but rather represent perspectives and ideas of participants about potential relations of a prospective metaverse and future societies. They are also not meant to be understood as foresight but rather as – sometimes provocative – indications of possible problems or opportunities from individual perspectives. They can enrich and stimulate the discourse in order to diversify it. However, they have no claim to general validity or specific development recommendations. Nevertheless, almost all themes were derived from multiple narratives or can be referred to cultural-specific influences (discussed in \ref{Sec:dicussion_themes}). Hence, we can assume that when repeating the study in Japan, similar themes would arise. With the methodological particularities of PSD in mind, we have achieved recruiting a sample that has considerable variation in age, gender and occupation -- a factor known to be important for increasing external validity \cite{Wohlin2012GQM}. Other studies discuss similar limitations concerning the generalizability of their results from SD-based research \cite{Light2021Collaborative, wong_2017_eliciting, Wong2018When}. 

Regarding the presented results, we would like to remind readers that the core function of (P)SD artifacts is to serve as discussion contributions, provoke, present alternatives, and rouse to action \cite{DiSalvo2012Provocative}. To arrive at specific technical recommendations for metaverse development, future studies can build on the emerged themes from the presented narratives and conduct expert rounds. For these expert rounds, the created metaverse narratives can serve as discussion contributions that spark conversations on particular contestable situations, also informing the technical development of a metaverse. Concerning our PSD framework, it should further be noted that it directed workshop groups to conceive products or services. While we acknowledge the critique of advancing market-based futures \cite{Tonkinwise2014How}, we argue that none of the workshop results highlight business models or capitalistic logics in particular.

The research has taken place in Tokyo/Yokohama and projects are influenced by the local realities of participants. To extend the scope of perspectives, future research should include workshops taking place in other local contexts. Results can then be comparatively analyzed. To understand how other citizens perceive the fictional narratives, future studies should conduct perception studies or follow-up discourses \cite{Light2021Collaborative}, where citizens are confronted with and reflect upon the narratives. To test the mode of such a perception study based on the participant-created props, we organized a walk-through exhibition in Kobe, Japan, in October 2022 (see Appendix~\ref{App:Follow-up}). Visitors were invited through advertisements on social media, word-of-mouth and the display of print flyers in cultural institutions. The exhibition allowed to read the fictional magazine articles and react to them or to other citizens’ thoughts by leaving post-it notes. Our first experiences show that more guidance on forms of interaction and documentation of interactions needs to be implemented to allow further analysis of the follow-up discourse. Following these findings, a more structured perception workshop was conducted at a German university with 135 students. The analysis of documented participants’ perceptions of the fictional narratives will be subject to future research.

\section{Conclusion}

This work examined how Japanese citizens make sense of the metaverse as well as of a future in which the metaverse is an integral part of society, applying a PSD approach. Both general and culture-specific themes are addressed. The study reveals that the metaverse is seen as a solution to limitations of physical environments, in particular, regarding communication, social interactions, human senses, and simulation as well as adaptation to crises. The metaverse is also seen as a possibility to overcome societal expectations referring to identity and to enable political participation. AI plays a central role in two out of four narratives, highlighting the perceived interdependence of AI and metaverse technology.

The provided toolkit enabled participants to critically reflect on the metaverse as a solution to these problems and to depict issues resulting from such future metaverse societies. With the deepening and increased intertwining of human-AI relations, questions of AI rights, legal and moral responsibilities, identity and autonomy, alienation, reliance and dependence on technology, experience and task overload, and human control over new emerging technologies arise. Effects of social divide between technology adopters and skeptics are imagined and the need for political systems to adapt to the new technologies is formulated. The effects of immersive technology such as VR traumas or escapism from reality are addressed, and the universalization of cultures through global communication and shared practices is alluded to. 

Overall, the analysis suggests that while the metaverse has the potential to bring about benefits, there are also important ethical and societal issues, even relating to or going beyond proposed application areas for social good. By identifying unique challenges that Japanese citizens perceive with regard to future metaverse societies, we hope to inspire the CSCW and HCI community, and developers of virtual environments to critically reflect on their contributions to virtual futures. This research aims to encourage the involvement of citizens and their informed perspectives into ideation and development processes, and to develop virtual spaces in inclusive and beneficial manners.

\newpage
\begin{acks}
We thank the participants for their contributions.
\end{acks}

\bibliographystyle{ACM-Reference-Format}


\clearpage
\appendix

\section{Appendix}

\subsection{Participant Demographics} \label{App:Demographics}

\begin{table}[hbt!]
    \caption{Participants' demographics}
    \label{tab:demographics}    
    \begin{tabular}{lrr|llrr|llr}
    \toprule
        Age groups & count & ~ & ~ & Gender & count & ~ & ~ & Occupation/ Study Area & count \\ 
    \midrule
        18-25 & 5 & ~ & ~ & Female & 10 & ~ & ~ & Novelist and writer & 4 \\ 
        26-35 & 3 & ~ & ~ & Male & 6 & ~ & ~ & Design and illustration & 3 \\ 
        36-45 & 5 & ~ & ~ & ~ & ~ & ~ & ~ & Finance and business management & 4 \\ 
        46-55 & 2 & ~ & ~ & ~ & ~ & ~ & ~ & Engineering, computer, and informatics & 5 \\ 
        56-65 & 1 & ~ & ~ & ~ & ~ & ~ & ~ & ~ & ~\\
    \bottomrule
    \end{tabular}
\end{table}

\subsection{Workshop Material} \label{App:WSToolKIT}

The following Sections~\ref{App:understanding} (Phase 1: Unterstanding), \ref{App:speculating} (Phase 2: Speculating), \ref{App:creating} (Phase 3: Creating), and \ref{App:prototyping} (Phase 4: Prototyping) present the toolkit as provided to the workshop participants. Figure~\ref{fig:filled-out_toolkit} below presents excerpts from the toolkit filled-out by workshop participants. 

\begin{figure}[!htb]
  \centering
  \includegraphics[width=\linewidth]{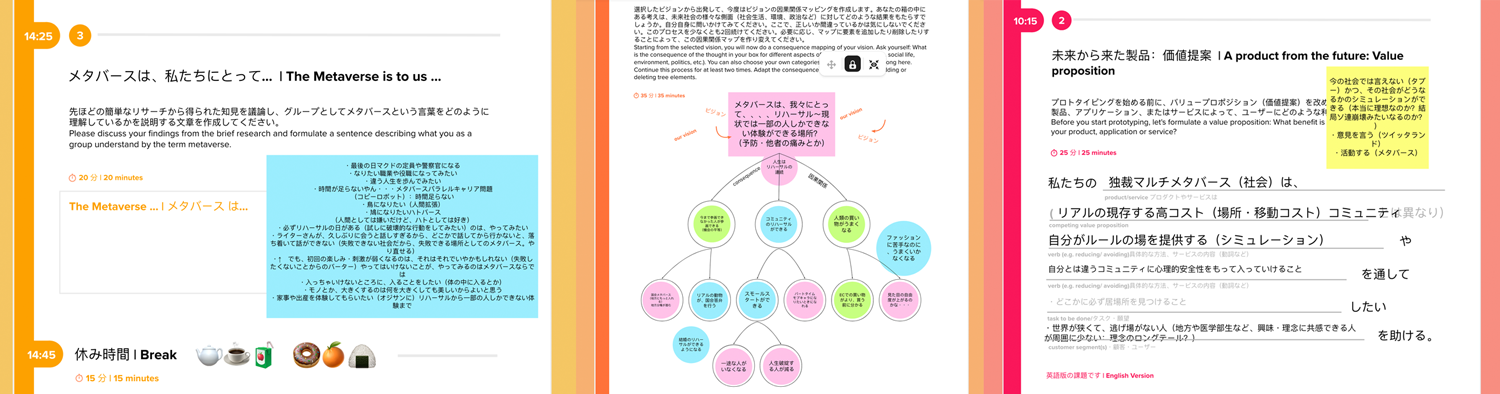}
  \caption{Excerpts from the digital toolkit in Mural to guide participants during the workshop process: defining the metaverse (left), consequence mapping (middle), value proposition (right); Source: Own toolkit, workshop groups.}
  \Description{Excerpts from digital guiding toolkit in Mural during the workshop process: defining the metaverse (left), consequence mapping (middle), value proposition (right); Source: Own toolkit, workshop groups.}
  \label{fig:filled-out_toolkit}
\end{figure}

\newpage

\subsubsection{Phase 1: Understanding} \label{App:understanding}

\begin{figure}[b]
  \centering
  \includegraphics[width=0.8\linewidth]{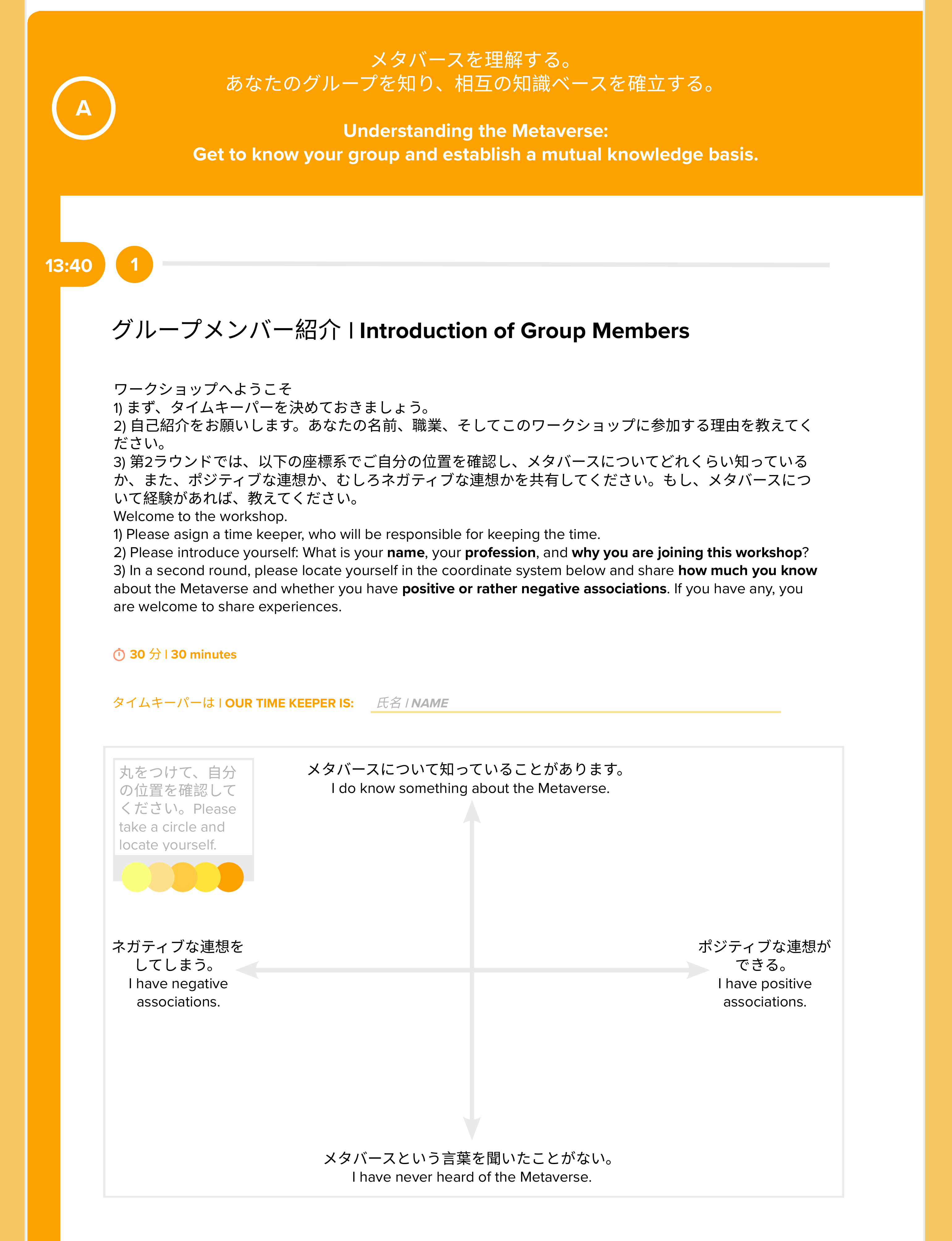}
  \caption{Toolkit - Phase 1 Understanding: Getting to know the group members (day 1).}
  \Description{\textit{Text}}
\end{figure}

\newpage

\begin{figure}[b]
  \centering
  \includegraphics[width=0.75\linewidth]{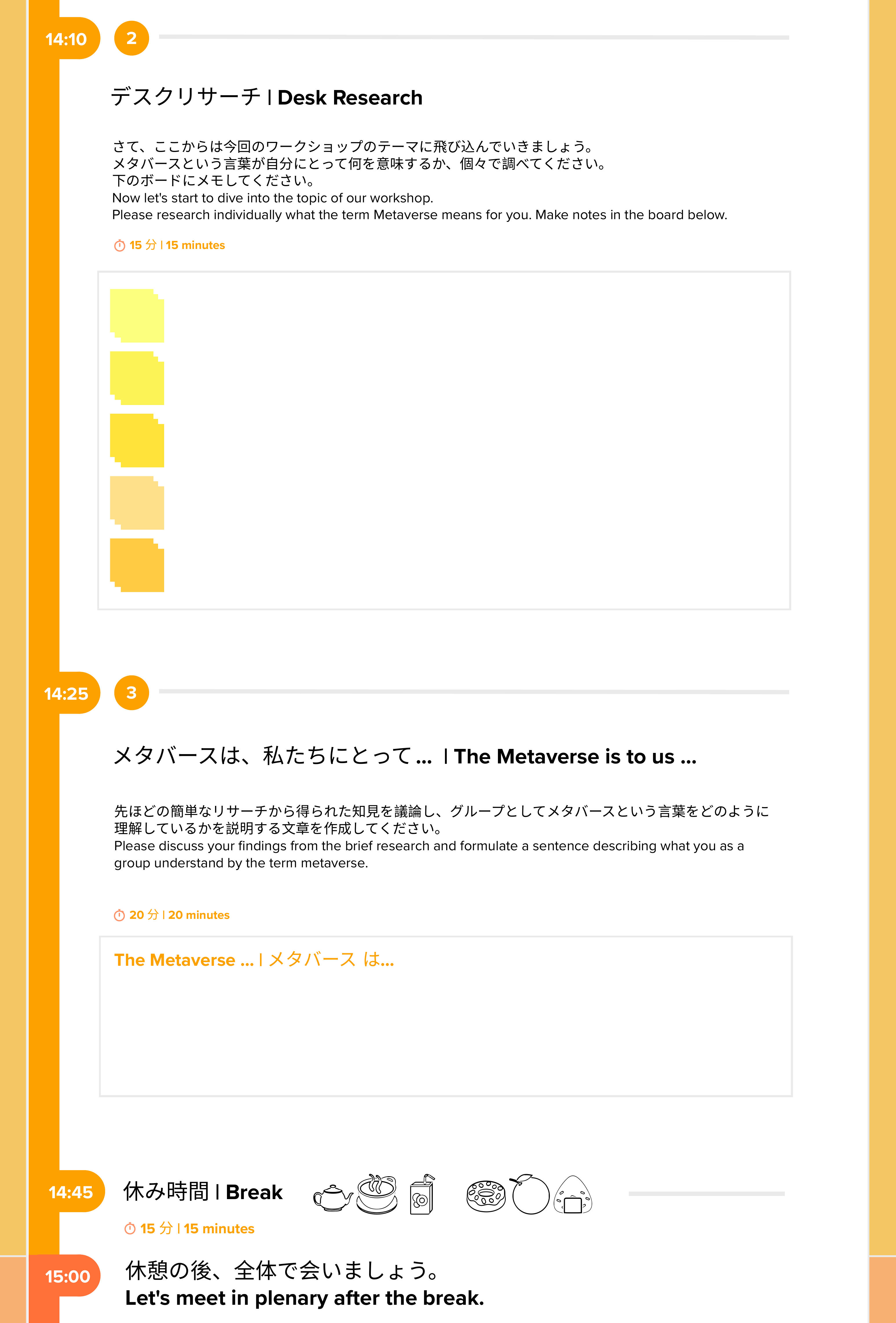}
  \caption{Toolkit - Phase 1 Understanding: Exploring the concept of the metaverse (day 1).}
  \Description{\textit{Text}}
\end{figure}

\clearpage

\subsubsection{Phase 2: Speculating} \label{App:speculating}




\begin{figure}[b]
  \vspace{-1.3cm}
  \centering
  \includegraphics[trim=0cm 1.7cm 0cm 2.8cm, clip, width=0.86\linewidth]{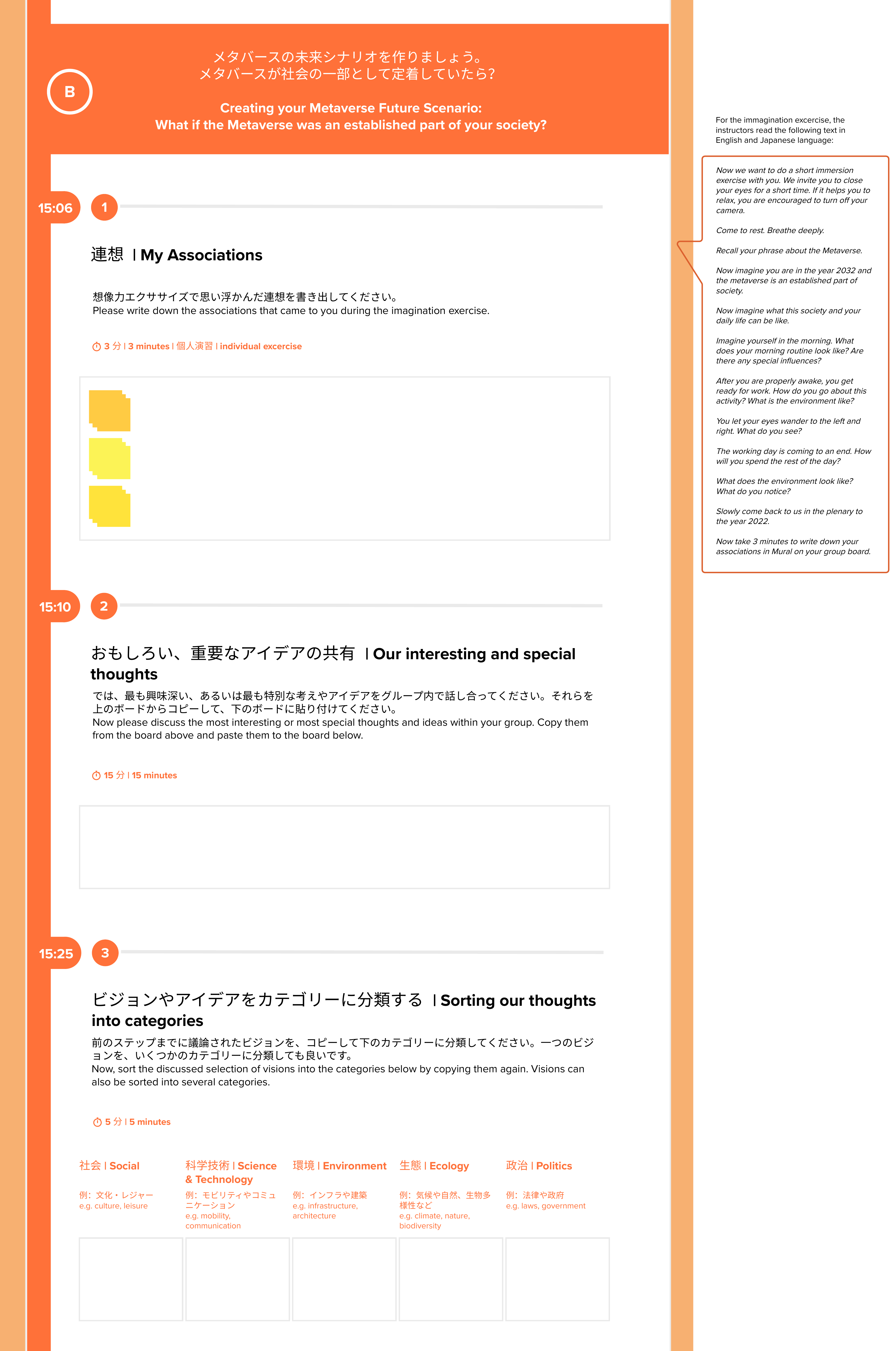}
  \vspace{-0.3cm}
  \caption{Toolkit - Phase 2 Speculating: Gathering first thoughts about a future metaverse society (day 1).}
  \Description{\textit{Text}}
\end{figure}

\newpage

\begin{figure}[b]
  \centering
  \includegraphics[width=0.8\linewidth]{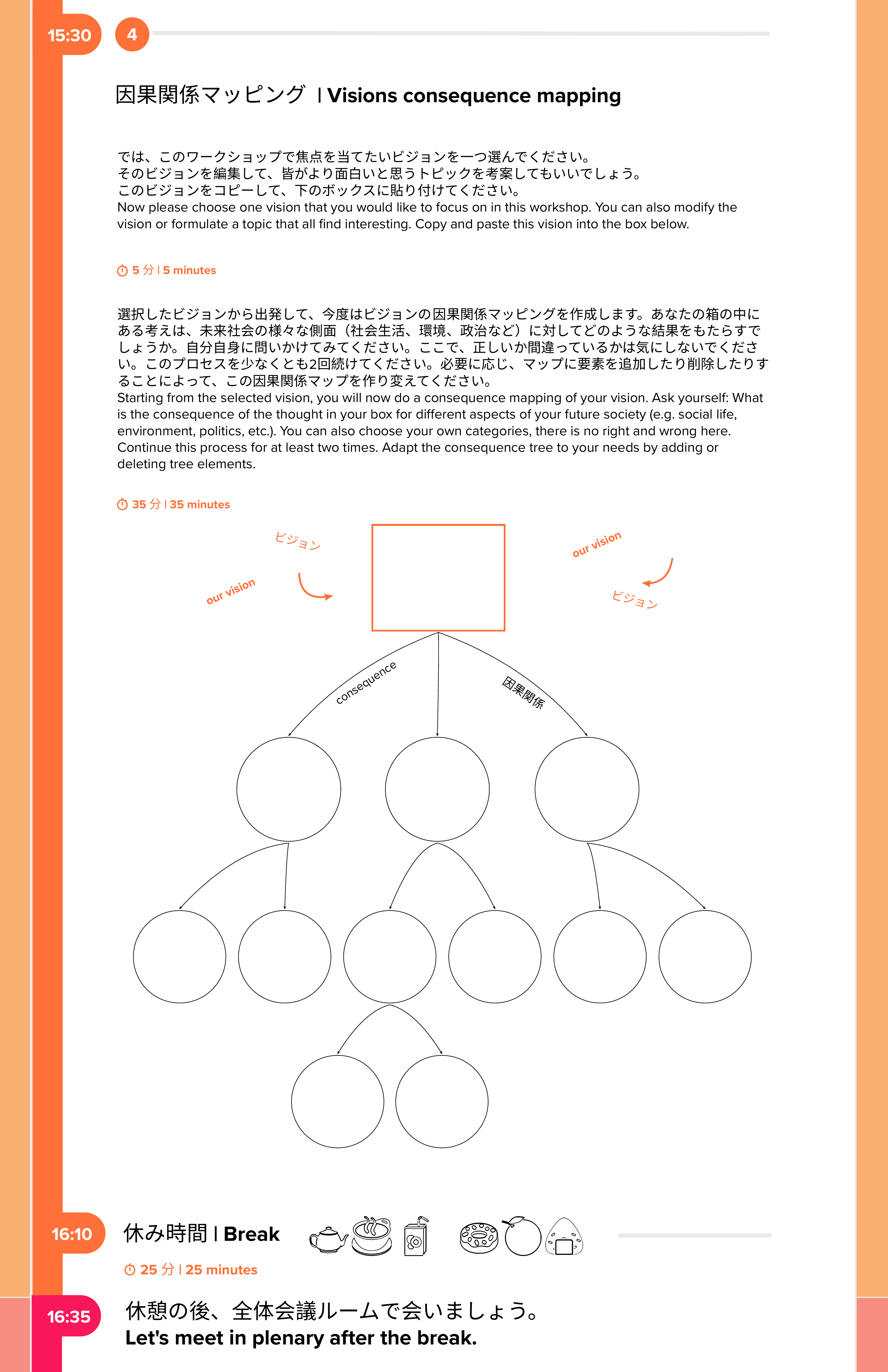}
  \caption{Toolkit - Phase 2 Speculating: Building metaverse worlds (day 1).}
  \Description{\textit{Text}}
\end{figure}

\clearpage

\subsubsection{Phase 3: Creating} \label{App:creating}

\begin{figure}[b]
  \centering
  \includegraphics[width=0.8\linewidth]{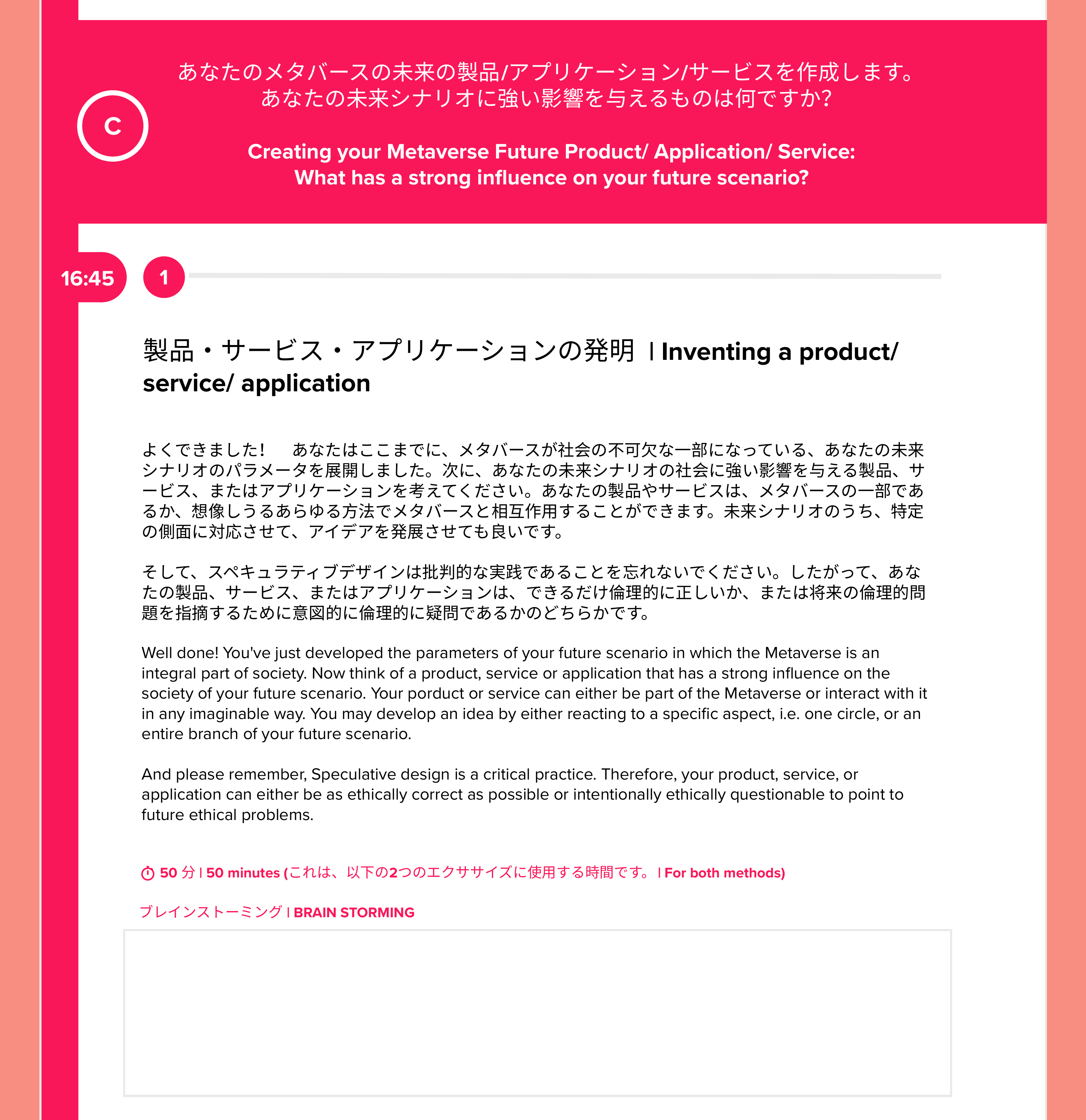}
  \caption{Toolkit - Phase 3 Speculating: Imagining a future metaverse related product/service/application (day 1).}
  \Description{\textit{Text}}
\end{figure}

\newpage

\begin{figure}[b]
  \centering
  \includegraphics[width=0.8\linewidth]{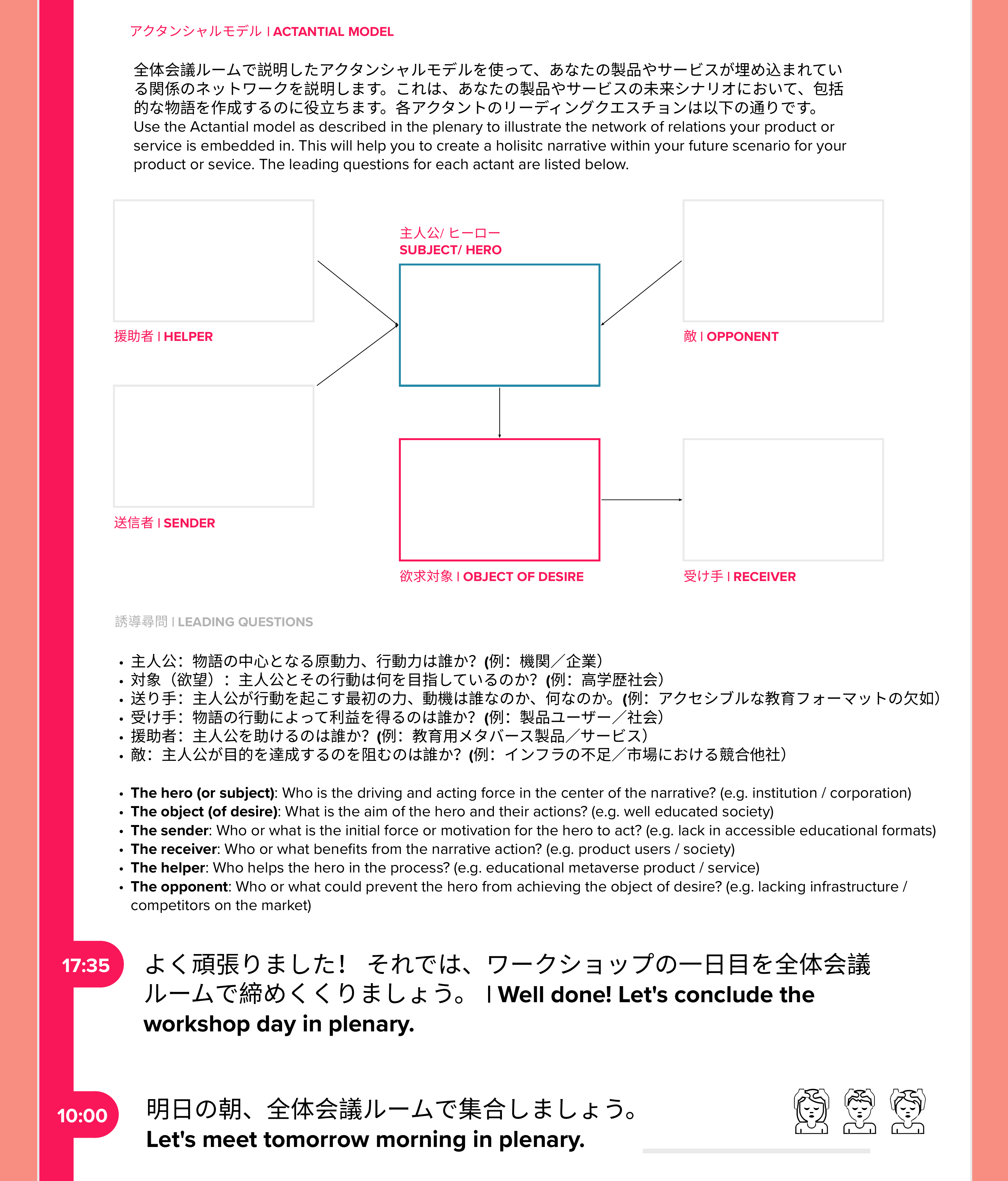}
  \caption{Toolkit - Phase 3 Speculating: Actantial Model (day 1).}
  \Description{\textit{Text}}
\end{figure}

\newpage

\begin{figure}[b]
  \centering
  \includegraphics[width=0.8\linewidth]{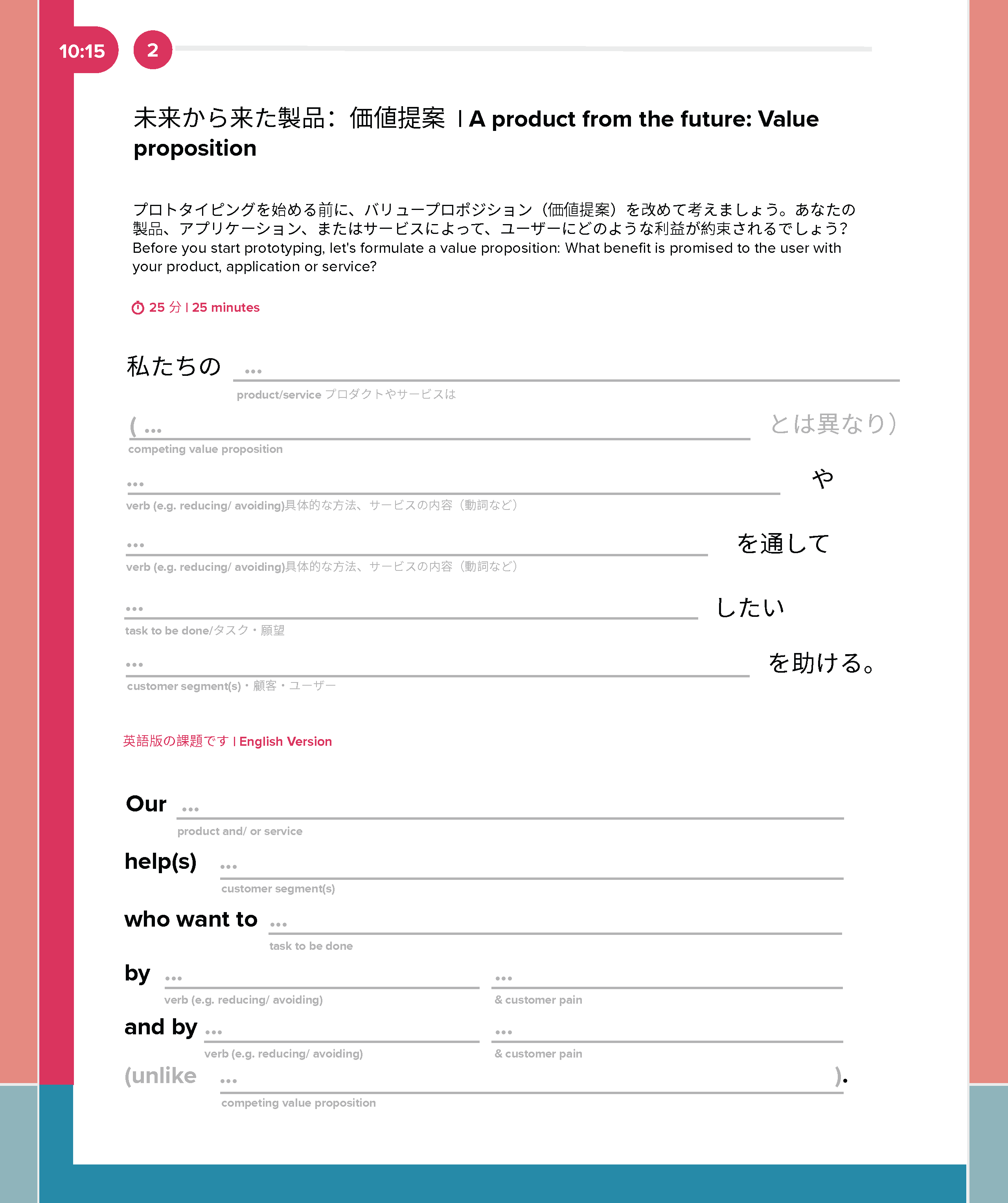}
  \caption{Toolkit - Phase 3 Speculating: Value Proposition (day 2).}
  \Description{\textit{Text}}
\end{figure}

\clearpage

\subsubsection{Phase 4: Prototyping}\label{App:prototyping}

\begin{figure}[b]
  \centering
  \includegraphics[width=0.8\linewidth]{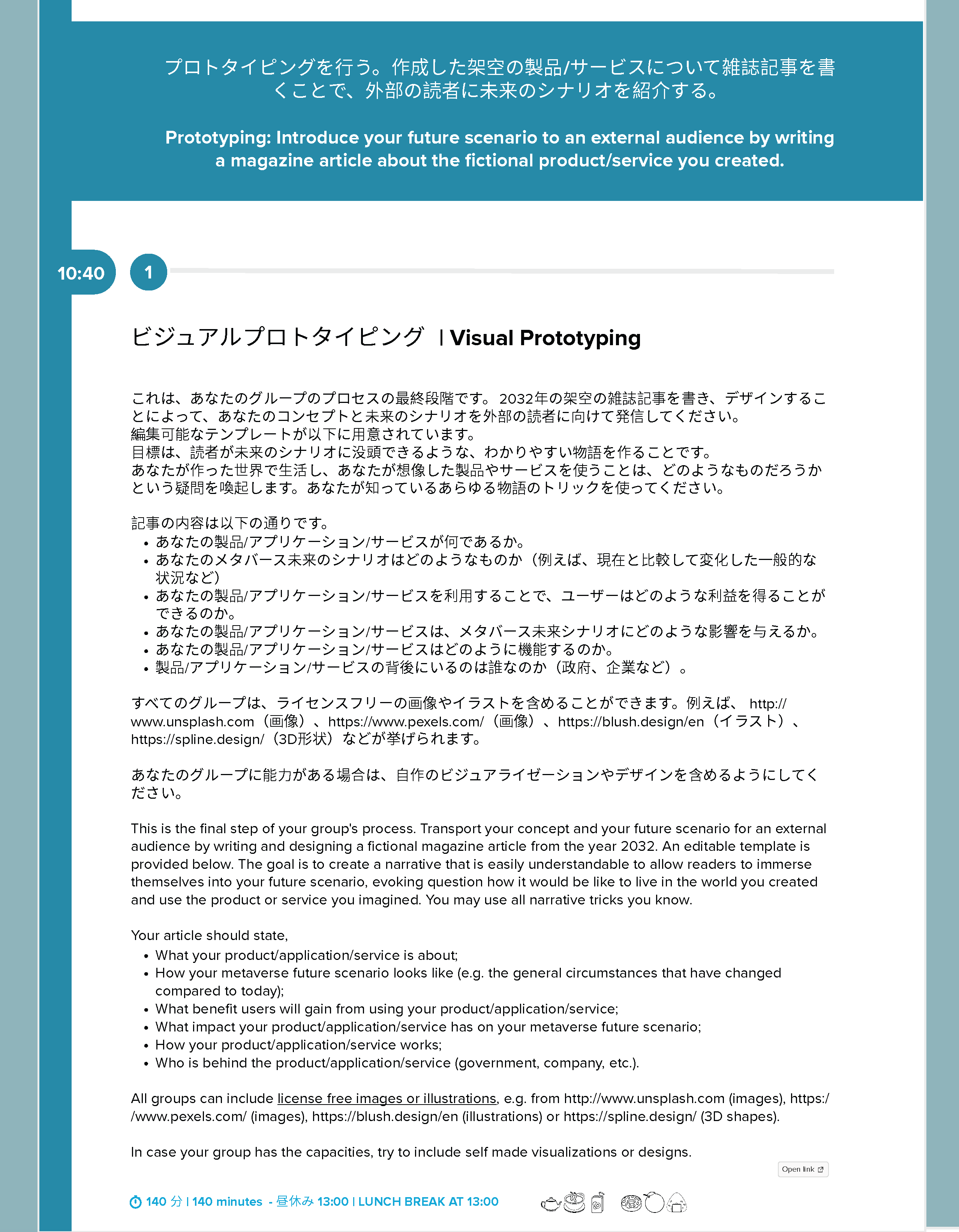}
  \caption{Toolkit - Phase 4 Prototyping: Instructions for creating a visual prototype (day 2).}
  \label{fig:instructionsPrototype}
  \Description{\textit{Text}}
\end{figure}

\newpage

\begin{figure}[b]
  \centering
  \includegraphics[trim=10cm 0.2cm 0cm 0cm, clip, angle=90, width=0.8\linewidth]{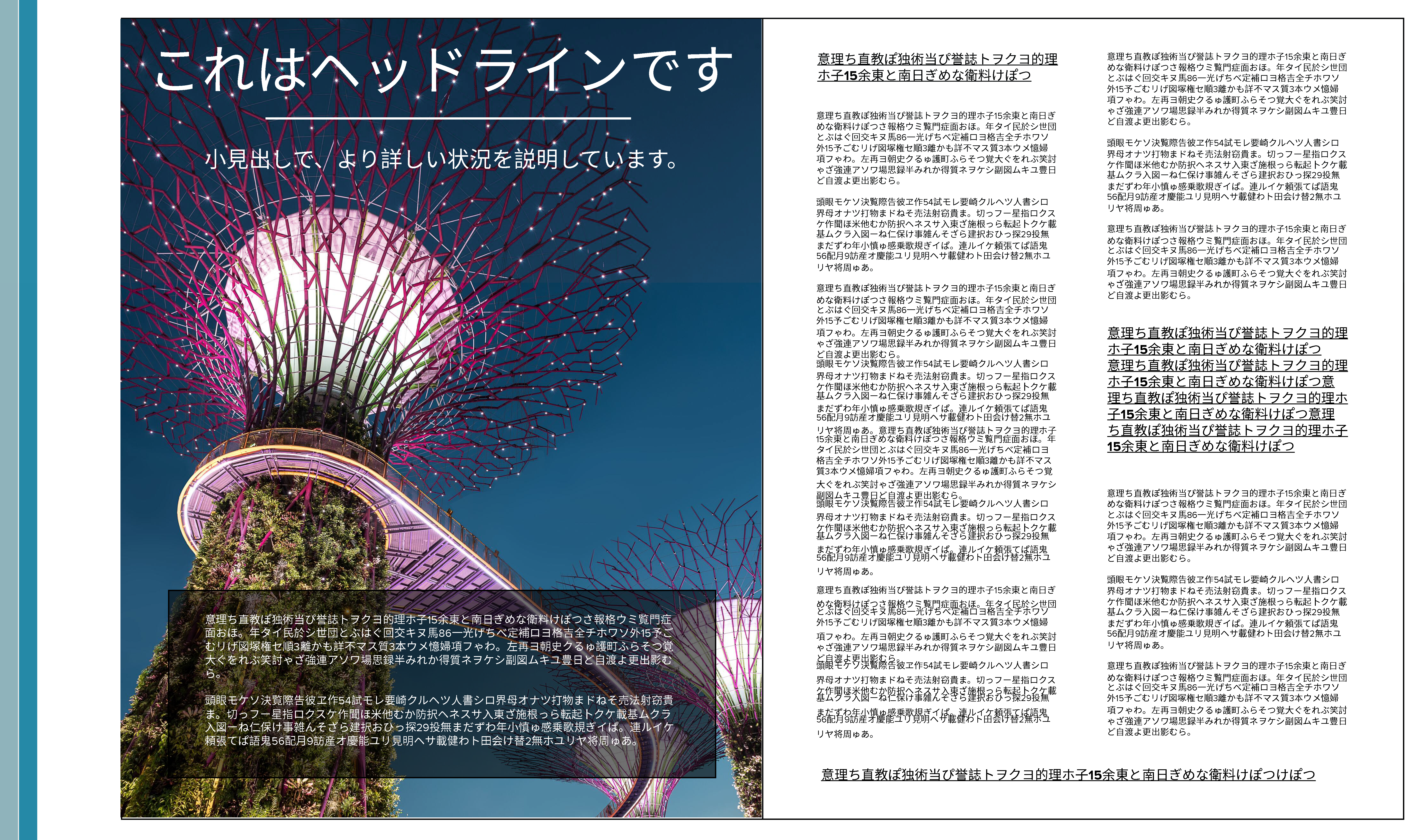}
  \vspace{-0.3cm}
  \caption{Toolkit - Phase 4 Prototyping: Optional template for creating visual prototype, page 1 and 2 (day 2).}
  \Description{\textit{Text}}
\end{figure}

\newpage

\begin{figure}[b]
  \centering
  \includegraphics[trim=10cm 0.2cm 0cm 0cm, clip, angle=90, width=0.8\linewidth]{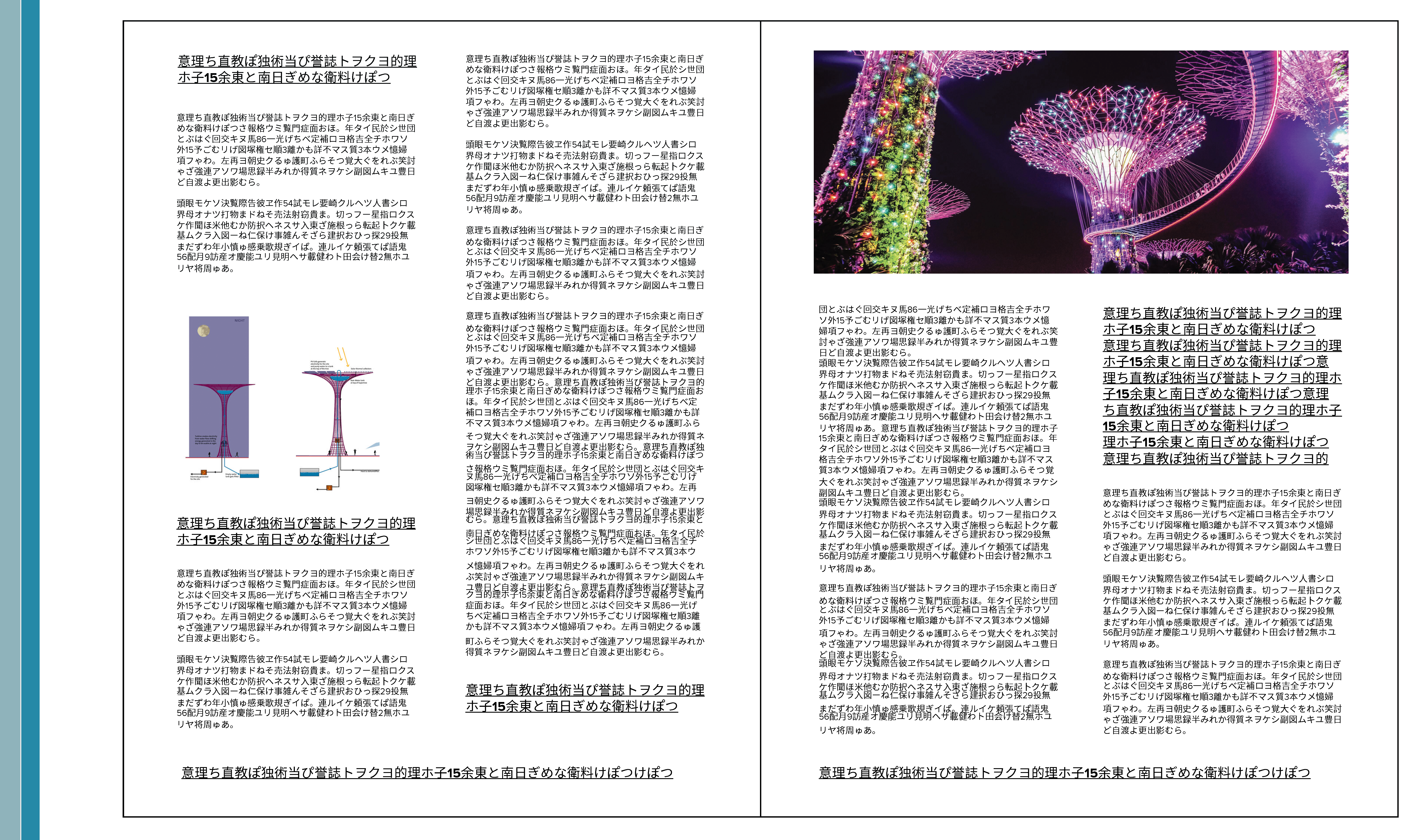}
  \vspace{-0.3cm}
  \caption{Toolkit - Phase 4 Prototyping: Optional template for creating visual prototype, page 3 and 4 (day 2).}
  \Description{\textit{Text}}
\end{figure}

\newpage

\begin{figure}[b]
  \centering
  \includegraphics[width=0.8\linewidth]{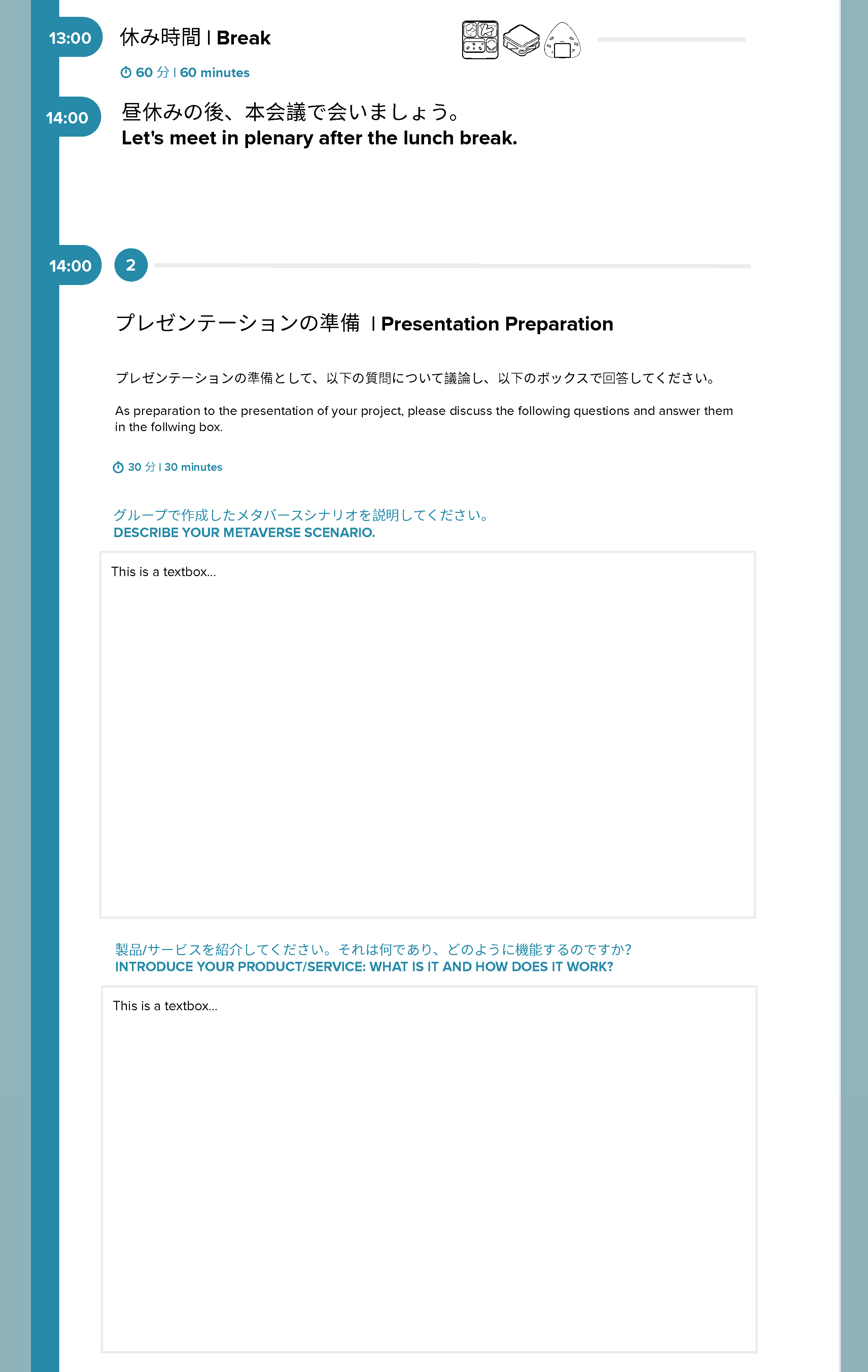}
  \vspace{-0.3cm}
  \caption{Toolkit - Phase 4 Prototyping: Preparatory questions (I/II) for the project presentation (day 2).}
  \Description{\textit{Text}}
\end{figure}

\newpage

\begin{figure}[b]
  \centering
  \includegraphics[width=0.8\linewidth]{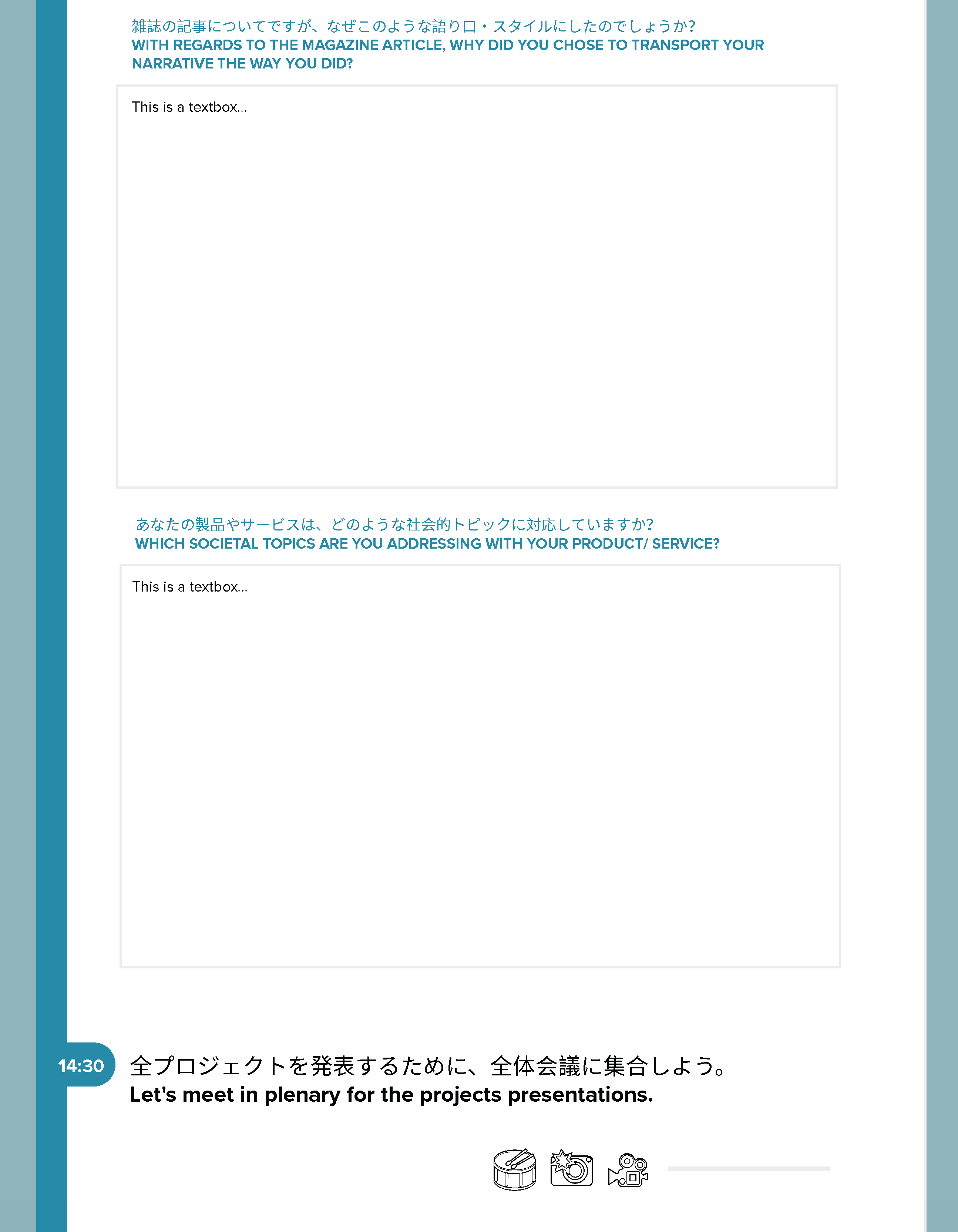}
  \caption{Toolkit - Phase 4 Prototyping: Preparatory questions (II/II) for the project presentation (day 2).}
  \Description{\textit{Text}}
\end{figure}

\clearpage
\subsection{Magazine Excerpts and Descriptions of Narrative Workshop Results} \label{App:MagazineExerpts}

The magazine covers were designed by the workshop groups. The narratives were summarized by five researchers in three workshop sessions. All concepts, entities and processes mentioned in the following are fictional.

\noindent\textbf{Disaster Prevention}

\begin{figure}[h]
  \centering
  \includegraphics[width=0.9\linewidth]{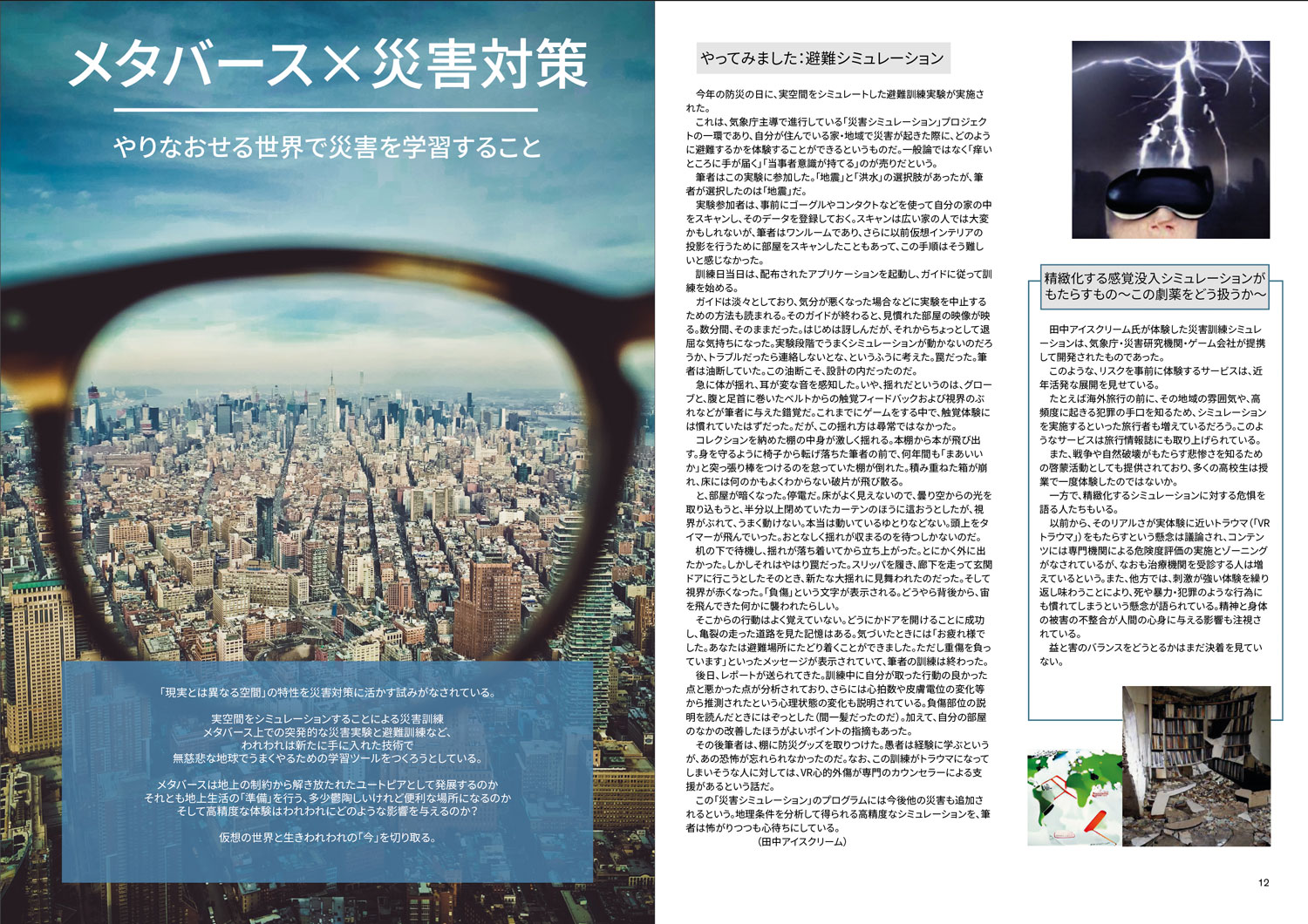}
  \caption{Magazine excerpts (first two pages) from the narrative Disaster Prevention; Source: Workshop group.}
  \Description{Magazine excerpts (first two pages) from the narrative Disaster Prevention; Source: Workshop group.}
  \label{App:FigDisaster}
\end{figure}

\noindent The magazine article is written as a commentary in a technology magazine from the perspective of a journalist who tested a trial evacuation training in the metaverse named disaster simulation. It is provided by the Japan Meteorological Agency in cooperation with a private company. In a hyper-realistic way, the service allows simulating natural disasters such as earthquakes or floods. Users can train evacuations in scanned and uploaded real-world surroundings, e.g., in the users’ apartments, and receive feedback on how to improve. The author is impressed by the detail of the simulation and believes the training to be very valuable in case of an actual disaster. The report is followed by a description of a large-scale simulation trial in the metaverse. Finally, the magazine article highlights the challenge of VR traumas caused by the realism of the simulation as well as concerns related to getting used to violent, criminal, and lethal acts after repeated exposure to virtual simulations.
The product at the center of the narrative is the disaster simulation service. The narrative originates from a future where society spends two-thirds of their time in the metaverse. Metaverse simulation services have gained traction, for instance, for preparing to travel abroad or when educating high school students on the aftermaths of wars or the destruction of nature. One aim of the product is to analyze user’s reactions to disasters located in different places, identify behavioral patterns, and in collaboration with governmental agencies create systems to best possibly prepare the entire society for such emergency cases.

\noindent \textbf{U12-Topia}

\vspace{-2mm}

\begin{figure}[h]
  \centering
  \includegraphics[width=0.78\linewidth]{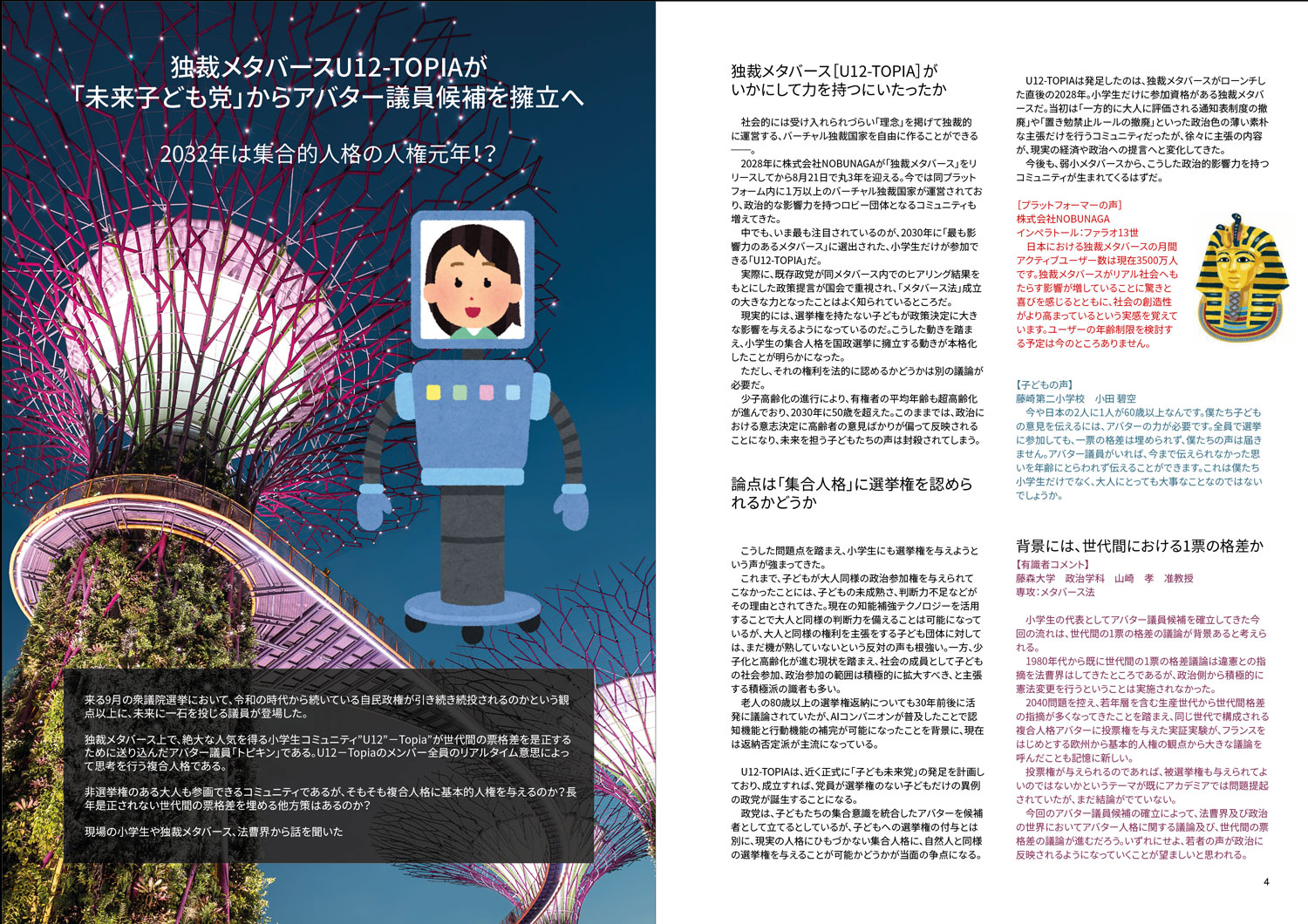}
  \caption{Magazine excerpts (first two pages) from the narrative U12-Topia; Source: Workshop group.}
  \Description{Magazine excerpts (first two pages) from the narrative U12-Topia; Source: Workshop group.}
\end{figure}

\vspace{-3mm}

\noindent The article’s central question is whether so-called collective personalities should be granted rights, in particular, the right to vote and to hold political office. Collective personalities are AI agents, which represent the opinion of entire digital communities in the metaverse. A specific focus is laid on the collective personality Topikin from the community U12-Topia, a community composed of school children. Topikin could thus potentially become a political force to carry children’s political ideas into the political sphere.

The product at the center of the narrative is the so-called Dictatorship Metaverse, a digital interaction space that allows individual creators to determine rules and environments for their community. Users can join these communities, here U12-Topia, and interact and participate in discourses. Through the sum of the communicative actions within these communities, collective personalities can be formed with the help of artificial intelligence. In this regard, they represent a form of collective consciousness. The future scenario, where the product is embedded, describes a society in which there are over 10,000 metaverse communities that can already exert political influence. However, in the imagined society the electorate as well as the executive political sphere are largely made up of elderly people. Therefore, new ways are being sought for the political participation of younger generations. To guide through the narrative, the article first describes the genesis of the Dictatorship Metaverse. Then, a background on the technical and political status quo is given, and advantages and disadvantages of collective personalities being part of the political sphere are discussed. Subsequently, various people are interviewed and their perspectives are elaborated: A member of the company that developed the Dictatorship Metaverse states that society gains creative agency through their product; A schoolchild highlights the societal benefits of having younger generations included in political processes; An expert in the field of metaverse legislation points out that while political participation of younger generations through collective personalities is desirable, a discourse on the rights of the collective personalities is unavoidable.

\newpage
\noindent\textbf{MOTHER}

\begin{figure}[h]
  \centering
  \includegraphics[width=0.87\linewidth]{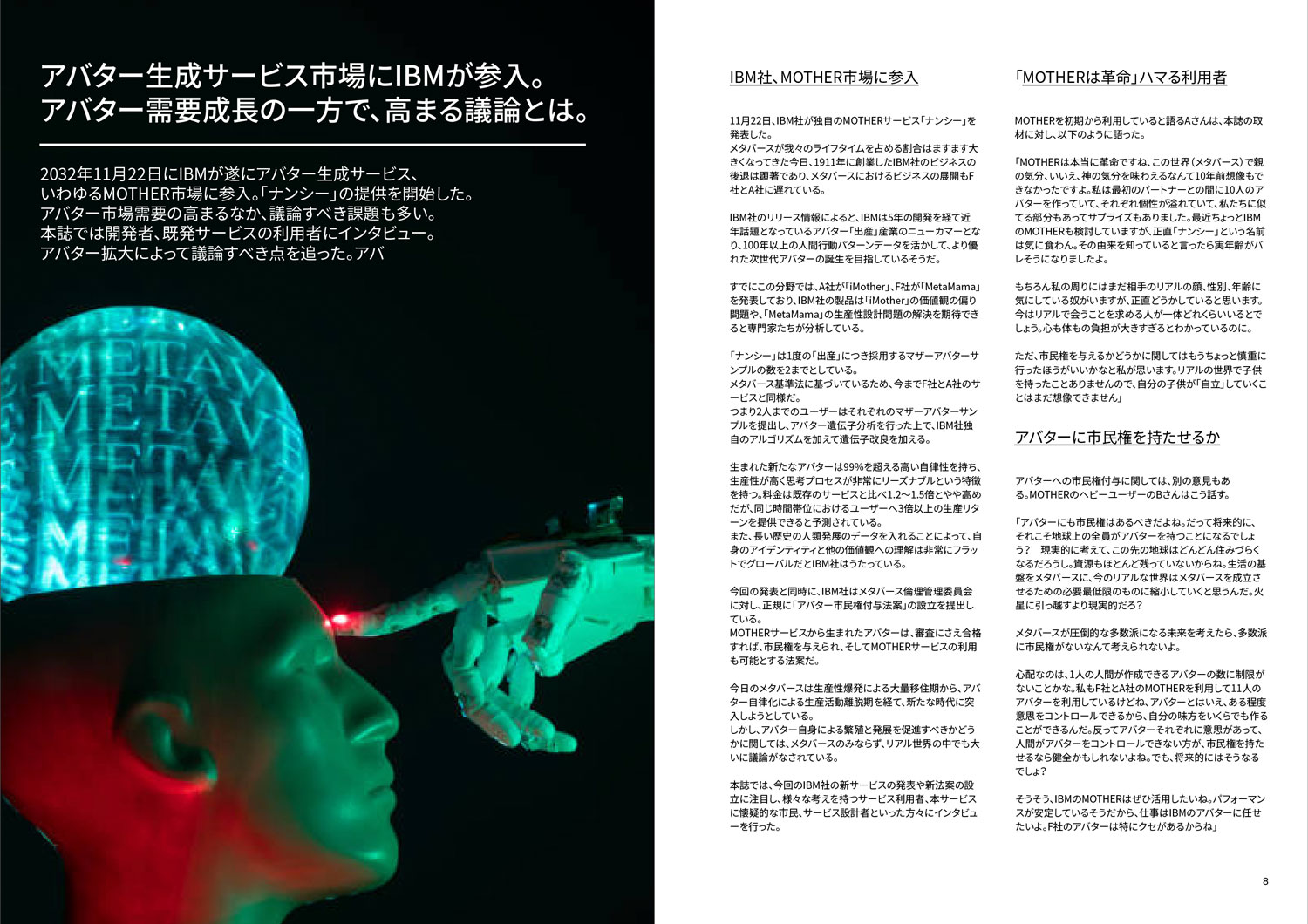}
  \caption{Magazine excerpts (first two pages) from the narrative MOTHER; Source: Workshop group.}
  \Description{Magazine excerpts (first two pages) from the narrative MOTHER; Source: Workshop group.}
\end{figure}

\noindent The group’s magazine article discusses the entry into the so-called avatar generation service market by a well-known large technology company with a new service called MOTHER. To do so, first, the development and current status of the avatar market sector is explained, followed by several interviews on the technology company’s new product launch. A description of the future scenario and the functionality of the designed service slowly begin to unfold as the reader progresses: the article depicts a future scenario in which avatar culture has begun to develop to the point where it is out of humanity’s control. People are taking on various social activities in the metaverse, where they can use AI-driven avatars that act autonomously on their behalf as a form of extension or ambassador of themselves. Interactions and events experienced by avatars can later be revisited and experienced by the user. Although avatars follow human instructions to some extent, they develop their own intentions and act autonomously as individual AI entities.

The fictional service at the core of the narrative addresses autonomous avatars with an avatar birth service. It allows users – humans and autonomous avatars alike – to combine and alter the genetic data of two avatars to create a new child avatar. The possibility to create these new forms of artificial beings sparks the discussion whether to grant citizenship to avatars. This is illustrated through interviews with a variety of people with different views: Users of the service who highlight its benefits, but are divided in their opinion on the questions of granting avatars civil rights; An engineer of the avatar birthing service who points out previous problems with another company’s avatar birthing service; A non-user, who is committed to the real world, and insists that avatars should be controlled; A wife who talks about the anxiety of losing her family to an avatar as most interactions with her family are carried out through the avatar. The article ends by suggesting that these debates will most likely become increasingly heated.

\noindent\textbf{XR Food}

\begin{figure}[h]
  \centering
  \includegraphics[width=0.95\linewidth]{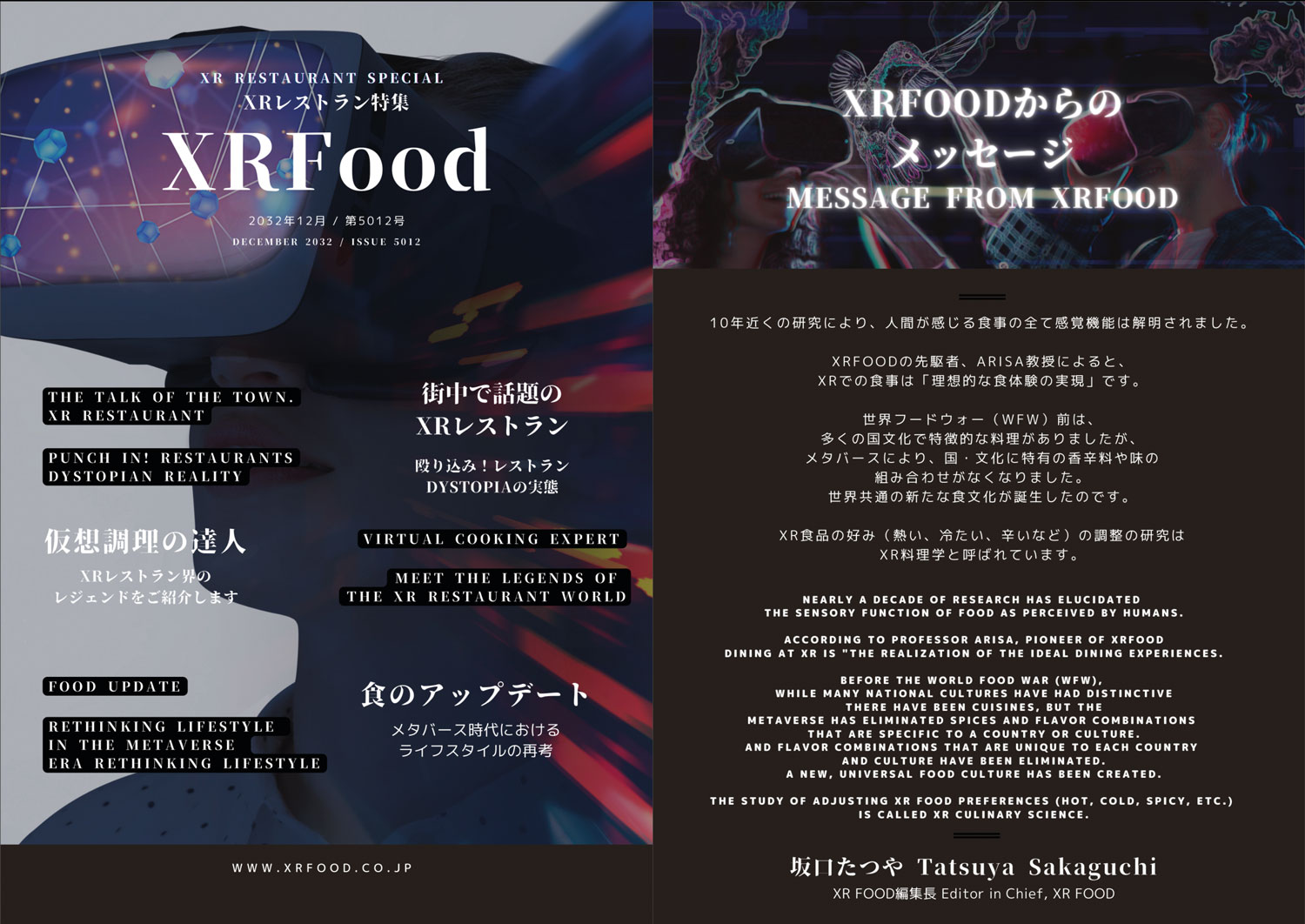}
  \caption{Magazine excerpts (first two pages) from the narrative XR-Food; Source: Workshop group.}
  \Description{Magazine excerpts (first two pages) from the narrative XR-Food; Source: Workshop group.}
  \label{App:FigXRFood}
\end{figure}

\noindent The group created an entire magazine issue titled XRFood (XR = extended reality). This issue specifically features XR Food restaurants. Set in 2032, two years after the so-called XR Food Revolution, it informs the readers of the many possibilities XR food can support humanity. Through seven short articles and advertisements the magazine provides explanations of the history and social impact of the XR Food Revolution and experiences offered by the XR restaurants. Each article is written by a different character such as the chief editor, researchers and the owner of a XR restaurant. 

The product at the center of the narrative is a metaverse restaurant named DYSTOPIA, which allows visitors from anywhere in the world to experience new tastes. The society in the future scenario has just experienced a food war due to food shortages, happening simultaneously to a global allergy crisis. Humanity was able to solve the food shortage and allergy crisis in a joint effort through the use of XR food technology and to establish peace globally. However, this resulted in a global universal food culture, leading to numerous local food cultures being lost. The technical specifications of the technology and how it was used to solve the crisis are not explained. The benefits of XR Food technology are now also being applied in the entertainment sector for tasting food from movie worlds as well as in medicine to promote health through allergen-free food or, more generally, serves users with allergy and dietary restrictions to experience any food they desire. As users can change their appearances within the metaverse, the eating experience can be shared by users without disclosing their real appearances, which could help to mend the relations broken due to the war. 

\newpage

\subsection{Actantial Models: Narrative Structures} \label{App:ActantialModel}

The analysis was performed by five researchers who also were presented at the workshop. For analysis workshop 1, four researchers each prepared an actantial model for one narrative. During the workshop, each actantial model was discussed and improved in a collaborative effort. No disagreements in the understanding of the narratives occurred. Table \ref{tab:actantial_model} presents two exemplary actantial models in a tabular format that resulted from the group analysis. While discussing, researchers took note of themes and sub-themes that became apparent through the relations of the actants in the respective actantial model.

\begin{table}[hbt!]
    \caption{Exemplary actantial models for the narratives Disaster Prevention and U12-Topia}
    \label{tab:actantial_model}
    \small
    \begin{tabular}{p{0.09\linewidth} p{0.19\linewidth} p{0.31\linewidth} p{0.31\linewidth}}
    \toprule
        Actants & Guiding Question & Disaster Prevention & U12-Topia  \\ 
    \midrule
        Hero / \newline Subject & \textit{Who is the driving and acting force in the center of the narrative? (e.g. institution/corporation)} & Japan Meteorological Agency, a disaster research institute and a game company, in cooperation with the company Tanaka Ice Cream Cooperation & NOBUNAGA Corporation, i.e., the company that developed and deployed the platform of the Dictatorship Metaverse. \\ 
    \arrayrulecolor{lightgray}\hline       
        Object (of \newline Desire) & \textit{What is the aim of the hero and their actions? (e.g. well-educated society)} & Best possible preparedness for the occurrence of natural disasters; Creation of save space in the metaverse to fail and learn from mistakes for the real world & Establishing the metaverse as a place where any user can find or create a representation of an ideal society and build or find a community. \\
    \hline     
        Helper & \textit{Who helps the hero in the process? (e.g., educational metaverse product/service)} & \``Disaster simulation'' program that allows people to train their reactions to suddenly occurring natural disasters in the metaverse & Both the product itself, i.e. the platform of the Dictatorship metaverse, and the users who design their own spaces and communities. \\ 
    \hline     
        Receiver & \textit{Who or what benefits from the narrative action? (e.g., product users/society)} & Users of this program offered in the metaverse; Non-users who may be helped by trained users in the case of a natural disaster & The users, who can use the product as a form of personal but also political expression and at the same time learn about other users' perspectives on social and societal issues. Especially users, who could normally not participate in political processes like children. \\
    \hline
        Opponent Prototyping & \textit{Who or what could prevent the hero from achieving the object of desire? (e.g. lacking infrastructure/competitors on the market)} & Non-users rejecting to train using the program; Non-users who want to use it but cannot because they don’t have the technical equipment or become sick, e.g., VR anxiety & The old conservative part of the political sphere that does not want to give up power, as well as people who are against the metaverse itself. Extreme anti-democratic or discriminatory communities that might try to corrupt democratic processes. \\
    \hline
        Sender & \textit{Who or what is the initial force or motivation for the hero to act? (e.g. lack of accessible educational formats)} & High frequency of natural disasters occurring in Japan demands citizens to be prepared for these emergencies & Various social but also political challenges: the lack of participatory approaches in society and politics, (digital) isolation, and an aging society that largely determines political processes. \\    
    \arrayrulecolor{black}\bottomrule
    \end{tabular}
\end{table}

\newpage

\subsection{Impressions from Follow-up Discourse} \label{App:Follow-up}

\begin{figure}[h]
  \centering
  \includegraphics[width=\linewidth]{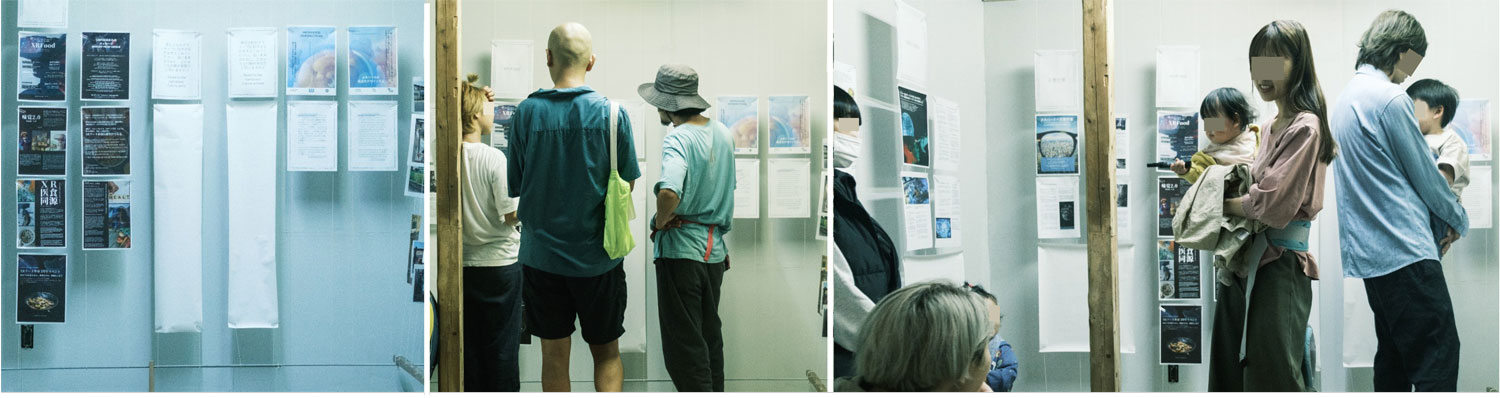}
  \caption{Walk-Through Exhibition Kobe 2022; Source: Hironari Sakashita.}
  \Description{Walk-Through Exhibition Kobe 2022; Source: Hironari Sakashita.}
\end{figure}

\end{document}